\documentclass[sigconf,10pt,printfolios=true]{acmart}
\renewcommand\footnotetextcopyrightpermission[1]{} 
\usepackage{tikz}
\usepackage[english]{babel}
\usepackage{blindtext}
\usepackage[T1]{fontenc}
\usepackage[utf8]{inputenc}
\usepackage{newtxmath,bm,courier}
\usepackage{xspace,enumitem,fancyhdr,fancyref,wrapfig}
\usepackage{amsmath,amsfonts,bm}
\usepackage{tabularx,tablefootnote,wrapfig,hyphenat}
\usepackage[labelformat=simple,skip=0pt]{subcaption}
\usepackage{xfrac,sparklines,hyperref}
\usepackage{algorithm,algpseudocode}

\usepackage{tabularx,wrapfig}
\usepackage[capitalize]{cleveref}

\setenumerate{itemsep=2pt,topsep=2pt,parsep=1pt,partopsep=0pt,leftmargin=1.5em}

\DeclareCaptionLabelSeparator{emdash}{--- }

\newcommand{\shortname}{CLCP}
\newcommand{\shortnameposs}{CLCP's}

\newcommand{\shortnames}{\shortnameposs}

\newcommand{\clcp}{CLCP}

\newcommand{\parabreak}{\vspace*{2.00ex minus 0.25ex}\noindent}
\newcommand{\parahead}[1]{\vspace*{1ex plus 0ex minus 0.25ex}\noindent{}{\bfseries #1}}
\let\oldReturn\Return
\renewcommand{\Return}{\State\oldReturn}
\acmDOI{}

\setcopyright{none} 

\begin{document}
\date{}
\title{\LARGE Cross-Link Channel Prediction for Massive IoT Networks}
\author{\large{\rm Kun Woo Cho$^1$, Marco Cominelli$^2$, Francesco Gringoli$^2$, Joerg Widmer$^3$, Kyle Jamieson$^1$}\\ Princeton Univ.$^1$, University of Brescia$^2$, IMDEA Networks$^3$}
\begin{abstract}
Tomorrow's massive-scale IoT sensor networks are poised to drive uplink traffic demand, especially in areas of dense deployment.
To meet this demand, however, network designers leverage tools that often require accurate estimates of Channel State Information (CSI), which incurs a high overhead and thus reduces network throughput.
Furthermore, the overhead generally scales with the number of clients, and so is of special concern in such massive IoT sensor networks.
While prior work has used transmissions over one frequency band to predict the channel of another frequency band on the same link, this paper takes the next step in the effort to reduce CSI overhead: predict the CSI of a nearby but distinct link.
We propose \emph{\textbf{Cross-Link Channel Prediction}} (\shortname{}), a technique that leverages multi-view representation learning to predict the channel response of a large number of users, thereby reducing channel estimation overhead further than previously possible.
\shortnames{} design is highly practical, 
exploiting channel estimates obtained from existing transmissions instead of 
dedicated channel sounding or extra pilot signals.
We have implemented \shortname{} for two different Wi-Fi versions, namely 802.11n and 802.11ax, the latter being the leading candidate for future IoT networks.
We evaluate \shortname{} in two large-scale indoor scenarios involving both line-of-sight and non-line-of-sight transmissions with up to $144$ different 802.11ax users.
Moreover, we measure its performance with four different channel bandwidths, from 20~MHz up to 160~MHz. 
Our results show that \shortname{}
provides a $2\times$ throughput gain over baseline 802.11ax and
a $30\%$ throughput gain over existing cross-band prediction algorithms. 

\end{abstract}

\maketitle
\fancyfoot{}
\setcopyright{none}

\section{Introduction}
\label{s:intro}

Today's wireless IoT sensor networks are changing, scaling up in spectral 
efficiency, radio count, and traffic volume as never seen before.
There are many compelling examples:~sensors in smart agriculture, 
warehouses, and smart\hyp{}city contexts
collect and transmit massive amounts of aggregate data, around the clock.
Networks of video cameras (\emph{e.g.}, for surveillance and in cashierless stores)
demand large amounts of uplink traffic in a more spatially\hyp{}concentrated pattern:
large retailers worldwide have recently introduced cashierless
stores that facilitate purchases via
hundreds of cameras streaming video to an edge server nearby, 
inferring the items the customer has placed into their basket as well
as tabulating each customer's total when they leave the store.
And in multiple rooms of the home, smart cameras, speakers, 
windows, and kitchen appliances stream their data continuously. 

\begin{figure}
\begin{subfigure}[b]{0.485\linewidth}
\includegraphics[width=.93\linewidth]{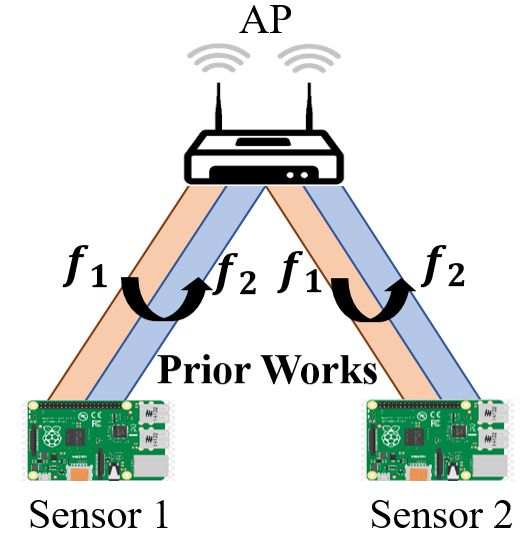}
\caption{Cross-band channel pred.}
\label{f:cbcp}
\end{subfigure}
\hfill
\begin{subfigure}[b]{0.485\linewidth}
\includegraphics[width=.93\linewidth]{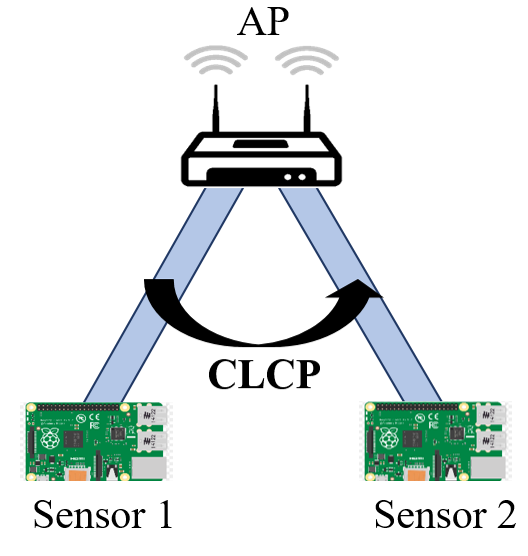}
\caption{Cross-link channel pred.}
\label{f:clcp}
\end{subfigure}
\caption{\textbf{\textit{Left:}} Previous work~\cite{R2F2-sigcomm16, bakshi2019fast} on cross-band channel prediction infers a 
downlink channel at frequency $f_{2}$ using the uplink channel
at frequency $f_{1}$ on the \emph{same} link. \textbf{\textit{Right:}} 
\shortname{} infers the channel to Sensor~2 using channel measurements from Sensor~1.}
\label{fig:intro_clcp}
\end{figure}

This sampling of the newest Internet of Things (IoT) 
applications highlights unprecedented demand for massive IoT device scale, 
together with ever\hyp{}increasing data rates.  
Sending and receiving data to these
devices benefits from advanced 
techniques such as Massive  Multi-User MIMO (MU-MIMO) and 
OFDMA-based channel allocation.  
The 802.11ax \cite{802.11ax-wc16} Wi-Fi standard, also known as \emph{Wi-Fi 6}, 
uses both these techniques for efficient transmission 
of large numbers of small frames, a good fit for IoT applications. 
In particular, OFDMA divides the frequency bandwidth into multiple subchannels, 
allowing simultaneous multi-user transmission.

While such techniques
achieve high spectral efficiency, they face a key challenge: they 
require estimates of channel state information (CSI), a process that hampers overall spectral efficiency. Measuring and propagating CSI to neighbors, in fact, scales with the product of the number of users, frequency bandwidth, antenna count, and frequency of measurement.
Highly\hyp{}dynamic, busy environments with human and vehicle
mobility further exacerbate these challenges, necessitating
more frequent CSI measurement. With densely deployed IoT devices, the overhead
of collecting CSI from all devices may thus deplete available radio resources~\cite{xie2013adaptive}.
While compressing CSI feedback
\cite{xie2013adaptive, cspy-mobicom13,bejarano2014mute} and\fshyp{}or 
leveraging channel reciprocity for implicit channel sounding \cite{R2F2-sigcomm16,bakshi2019fast,guerra2016opportunistic} 
reduces CSI overhead to some degree,
users still need to exchange compressed CSI
with the Access Point (AP) \cite{xie2013adaptive}, and implicit sounding 
relies on extremely
regular traffic patterns \cite{R2F2-sigcomm16, bakshi2019fast}, and so  
with increasing numbers of clients, AP antennas, and OFDM subcarriers, 
CSI overhead remains a significant burden.


\parabreak{}In this paper, we take a qualitatively different 
approach, inspired by the relative
regularity of IoT sensor traffic and the fact that a single wireless environment
is the determinant of nearby sensors' wireless channels.
While conventional wisdom holds that 
the channels of the nodes that are at least half a wavelength 
apart are independent due to link\hyp{}specific signal propagation 
paths \cite{tse-viswanath}, with enough background 
data and measurements of a wireless environment, we find that
it is possible to predict the CSI of a link that
has not been recently observed.
\cref{fig:intro_clcp} illustrates our high-level idea:
unlike previous works~\cite{R2F2-sigcomm16, bakshi2019fast} that
use CSI measurements at frequency $f_{1}$ to infer CSI at $f_{2}$ for a 
single link, our approach exploits the cross-correlation between
different links' wireless channels to leverage traffic on one sensor's link 
in order to predict the wireless channel of another.
We propose the \emph{Cross-Link Channel Prediction} (\shortname{}) method,
a wireless channel prediction technique that uses multiview representation 
machine learning to realize this vision.
We provide a head-to-head performance evaluation of our
approach against the OptML \cite{bakshi2019fast} and R2F2 \cite{R2F2-sigcomm16}
cross-band channel prediction methods in \Cref{s:eval}.

\begin{figure}
\centering
\includegraphics[width=1\linewidth]{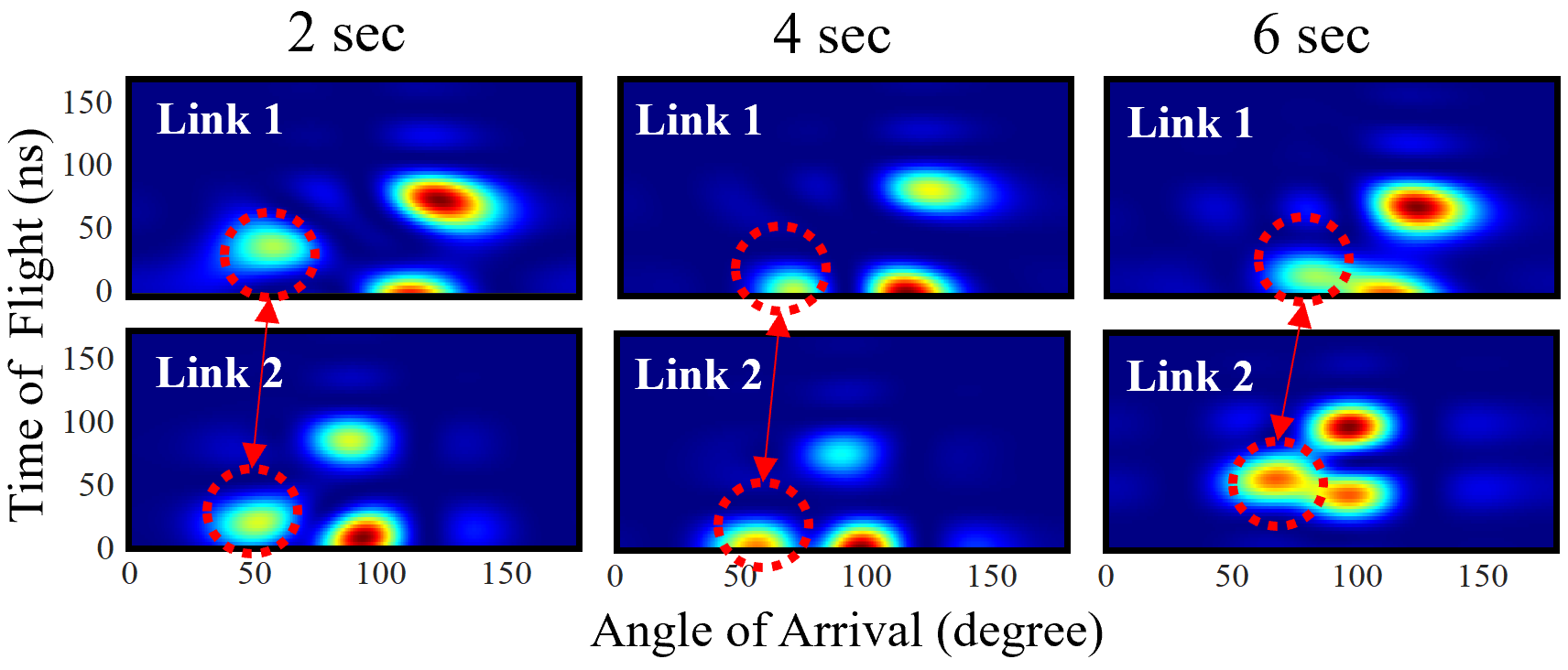}
\caption{\emph{\textbf{CLCP's mechanism:}} Time of 
flight and angle of arrival channel parameters for
two nearby IoT sensors (\textit{\textbf{upper}} and 
\textit{\textbf{lower}}, respectively). While each sensor (link) has a distinct set of 
static wireless paths, their parameters both indicate reflections
off the same moving object, highlighted in red dotted circles.
} 
\label{fig:passive_localization}
\end{figure}

To support our idea, we measure the wireless channels 
from two nearby sensors in the presence of a 
moving human. 
\Cref{fig:passive_localization} visualizes these channels using two wireless path parameters, 
Time-of-Flight (ToF) and Angle-of-Arrival (AoA). 
While Sensor~1's channel (upper pictures) 
is independent from Sensor~2's channel (lower pictures), ToF and AoA
from both reflect the same moving body in the environment, 
indicated by the dotted red circles, while
other major paths remain unchanged.
This suggests the existence of a function that correlates the occurrence of  wireless links of stationary sensors in the presence of moving reflectors,
as shown in \cref{f:feature_embedding}.

A \shortname{} AP can hence use uplink channels estimated from the nodes in the last transmission 
to predict a large number of unobserved wireless links.
In this way, the aggregated overhead no longer scales with the number of radios.
Finally, using the acquired CSIs, the AP schedules 
the uplink traffic.

Multiview learning for wireless channels faces several challenges,
which CLCP's design addresses. 
First, 
traffic patterns are not perfectly regular, and
a set of observed channels in latest transmission 
change at the time of prediction.
Hence, we cannot have a fixed set of observed channels as a model input.
Dynamic input data has been a big obstacle to multiview learning~\cite{zhu2020new, yang2018multi, wu2018multimodal} 
because it often leads to an explosion in the number of trainable parameters, 
making the learning process intractable.
Secondly, we often treat deep learning models as ``black boxes'' whose inner workings cannot be interpreted.
This is a critical issue because designers 
cannot differentiate whether the trained model truly works or is simply overfitting.
We summarize our key design points as follows:
\begin{enumerate}[label=\arabic*)]
\item \textbf{Low-overhead.}
Since the feedback overhead no longer scales with the number of radios,
\shortname{} incurs much lower overhead compared to prior work, when the number of wireless nodes is large. Hence, it improves the overall channel efficiency.

\item \textbf{Opportunistic.}
Unlike conventional approaches, \shortname{} does not need dedicated channel sounding or extra pilot signals for channel prediction. 
Instead, it exploits channel estimates obtained from existing transmissions on other links. 

\item \textbf{Low-power.}
802.11ax adopts a special power-saving mechanism 
in which the AP configures the timings of uplink transmissions
to increase the durations of nodes' sleep intervals.
%
By eliminating the need for channel sounding,
\shortname{} minimizes the frequency of wake-up and thus further reduces the power consumption. 
\item \textbf{Interpretable.}
Using \shortname{}, we visualize a fully trained feature representation and interpret it using the wireless path-parameters, ToF and AoA.
This allows network operators to understand \shortname{}'s learning mechanism and further help in modifying the system to match their needs.

\end{enumerate}
Our implementation and 
experimental evaluation using 802.11ax validate the effectiveness of
our system through microbenchmarks
as well as end-to-end performance.
End\hyp{}to\hyp{}end performance results show that \shortname{}
provides a $2\times$ throughput gain over baseline 802.11ax and
a $30\%$ throughput gain over R2F2
and a $30\%$  throughput gain over OptML 
in a 144-link testbed. 


\section{Primer: ML Background for \shortname{}}
The goal of \shortname{} is to correlate the occurrence of distinct wireless links given that they share \textit{some} views on the wireless environment. 
Specifically, we treat each channel reading
like a photo of an environment taken at a particular \textit{viewpoint}, and
combine multiple different views to form a joint representation of the environment. 
We then exploit
this representation to predict 
unobserved wireless CSI readings.
To do so, we must accomplish two tasks:
\begin{enumerate}
\item From the observed channels, we must discard radio-specific information and extract a feature representation that conveys information on the dynamics like moving reflectors. 
\item To synthesize unseen channels of a nearby radio, we need to integrate the extracted representation with radio-specific properties, including the signal paths and noises.
\end{enumerate}
However, 
radio-specific information and environment-specific information 
in the channel superimpose in channel readings and thus are not easily separable.
We exploit representation learning 
to capture a meaningful representation from 
a raw observation.
An encoder network of the representation learning model accomplishes 
the first task, and a decoder network of the model achieves the second task.
Before discussing the details of our \shortname{} design, we first 
provide some background on representation learning.

\begin{figure*}[t]
\includegraphics[width=\linewidth]{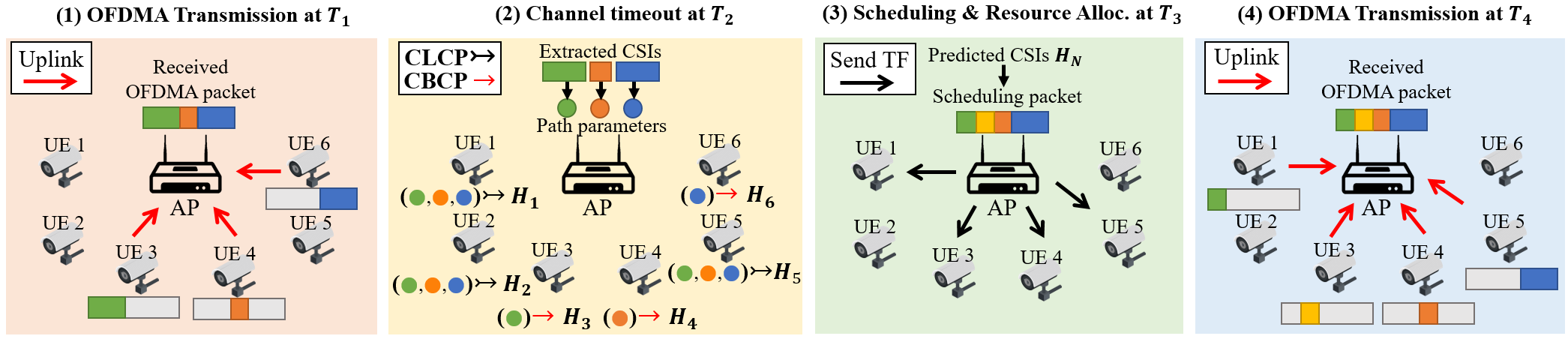}
\caption{\emph{\textbf{System overview for uplink transmission:}} 
(1) an AP receives uplink traffic from multiple users simultaneously.
(2) when channels become outdated, the AP extracts the path parameters of partial CSIs estimated from the latest OFDMA packet and predicts unobserved links using \shortname{} and unobserved bands using CBCP in a server; (3) Then, the AP schedules uplink traffic based on predicted CSIs and triggers the users. (4) Finally, the AP receives the scheduled packet.}
\label{fig:framework}
\end{figure*}

\parahead{Autoencoder.}
The \emph{autoencoder} 
learns lower\hyp{}dimensional representation $z$, that contains the information relevant for a given task. 
Specifically, an \textit{encoder} deep neural network (DNN) compresses the input data $h$ from the initial
space to the encoded space, also known as a latent space $z$, and 
a \textit{decoder} decompresses $z$ back to the data $\hat{p}$.  
However, it is not generalizable to new data and
prone to a severe overfitting.

\parahead{Variational Autoencoder (VAE).}
To enhance generalizability, the VAE \cite{kingma2013auto} integrates 
non\hyp{}determinism with the foregoing autoencoder.
In a nutshell, the VAE's encoder compresses data $h$ into a 
normal probability \emph{distribution} $z$, rather than discrete values. 
Then, it samples a point from $z$, 
and its decoder decompresses the sampled point back to the original data $\hat{h}$.
Mathematically, we represent a VAE model with a DNN $\theta$ as
$p_{\theta}(h,z)= p_{\theta}(z)p_{\theta}(h|z)$ 
where $p_{\theta}(z)$ is a prior (\textit{i.e.} Gaussian distribution)
and $p_{\theta}(h|z)$ is a decoder.
The goal of the training is to find a distribution that best describes the data.
To do so, it uses the
\emph{evidence lower bound} (ELBO):
\begin{equation}
    \mathrm{ELBO} 
        =\mathbb{E}_{q_{\phi}(z|h)}\left[\log\, p_{\theta}(h|z)\right]
        - D_{\mathrm{KL}}(q_{\phi}(z|h)||p(z))
        \label{eq:vae_term}
\end{equation}
where 
$q_{\phi}(z|h)$ is the encoder.
Its loss function consists of two terms where
the first term of the ELBO is 
the reconstruction error
and the second term is a regularization term.
This probabilistic approach has an advantage in predicting cross-link wireless channels
as it makes possible generalization beyond the training set~\cite{schonfeld2019generalized, klushyn2019increasing, bozkurt2019evaluating, bahuleyan2017variational}.



\parahead{Multiview Representation Learning.}
\textit{Multiview or multimodal
representation learning}~\cite{wu2018multimodal, pu2016variational, 
spurr2018cross, tsai2018learning}
has proven effective in 
capturing the correlation relationships of
information that comes as different \emph{modalities} (distinct data types or data sources).
For instance, different sets of photos of faces, each set having been 
taken at different angles, could each be considered different
modalities.
Such model learns correlations 
between different modalities 
and represents them jointly, 
such that the model can generate a (missing) instance of one modality 
given the others.
Like VAE, multiview learning encodes the primary view into a low-dimensional feature that contains useful information about a scene and decodes this feature, which describes the scene, into the secondary view.  

A more advanced form of multiview learning adopts multiple different views as input data and encodes them into a joint representation.  
By analyzing multiple information sources simultaneously,
we present an opportunity to learn a better, comprehensive feature representation.
For example, past work~\cite{sun2018multi} uses multiview learning to synthesize unseen image at an arbitrary angle given the images of a scene taken at various angles. 
Likewise, we treat each wireless link like a photo of a scene taken at a particular view-point.
We obtain the wireless link at many different view-points
and combine these views to form a joint representation of the channel environment. 
We then exploit this joint representation to predict the wireless link at unobserved view-points.

\section{Design}
\label{s:overview}

Our system operates in the sequence of steps shown in \Cref{fig:framework}:
first, an AP acquires \textit{buffer status report} (BSR) and \textit{channel 
state information} (CSI) from all clients. 
Then it schedules an uplink OFDMA packet based on obtained BSRs and CSIs and triggers the uplink transmission.
When acquired CSIs become outdated, 
the AP observes a set of channels from a latest OFDMA packet received and 
extracts the wireless path parameters from each channel. 
Then, the AP uses the path parameters to predict (1) the remaining bands of observed channels using \textit{cross-band channel prediction} (CBCP) and (2) a full-band channel information of unobserved links using \textit{cross-link channel prediction} (\clcp{}).   
Lastly, based on predicted CSIs, the AP runs a scheduling and resource allocation (SRA) algorithm and 
sends a trigger frame (TF) to initiate uplink transmission. 
The AP repeats the procedure whenever CSI readings are outdated.


\parahead{(1) Opportunistic Channel Observation.}
In OFDMA, the entire bandwidth is divided into multiple subchannels with 
each subchannel termed a resource unit (RU).
The AP assigns RUs to individual users, 
which allows one OFDMA packet to contain channel estimates from multiple users. 
We want to leverage channel information in already existing OFDMA transmissions to predict channels of a large number of users.
In Fig.~\ref{fig:framework}, 
user 3, 4, and 6 simultaneously transmit uplink signals in their dedicated RU, 
which sums up to a full band.
Once acquired CSIs time out, the AP estimates three subchannels from the received OFDMA packet
and uses them to predict not only the remaining subcarriers of observed links 
but also the full-band channel of unobserved users (\textit{i.e.,} user 1, 2, and 5). 
This way, we completely eliminate the need for channel sounding. 


\parahead{(2) Channel Prediction.}
When CSIs become outdated, the AP estimates the channels from the most recently received packet and
directly routes them to a backend server through an 
Ethernet connection. 
At the server side, the path parameters are extracted from partially observed channel estimates. 
These path parameters are then fed to \shortname{} for predicting all CSIs.
We further elaborate on \shortnames{} design in \cref{s:design}.

\parahead{(3) Scheduling and Resource Allocation (SRA).}
Lastly, the AP schedules the upcoming OFDMA transmission using predicted CSIs and 
a 11ax-oriented scheduling and resource allocation (SRA) algorithm.
We note that OFDMA scheduling requires 
a full-bandwidth channel estimate to allocate a valid combination of RUs of varying subcarrier sizes~\cite{wang-infocom17} and to find a proper modulation and coding 
scheme (MCS) index for each assigned RU.
Moreover, unlike 11n and 11ac, 11ax 
provides support for uplink MU-MIMO, 
which requires CSI to find an optimal set of 
users with low spatial channel correlation and an appropriate decoding precedence.
After computing the user-specific parameters required for uplink transmission,
the AP encloses them in a \textit{trigger frame} (TF) and 
broadcasts it as illustrated in Fig.~\ref{fig:framework}.

\parahead{(4) Uplink data transmission.} After receiving the TF, the corresponding users transmit data according to the TF.
\begin{figure}
    \includegraphics[width=0.95\linewidth]{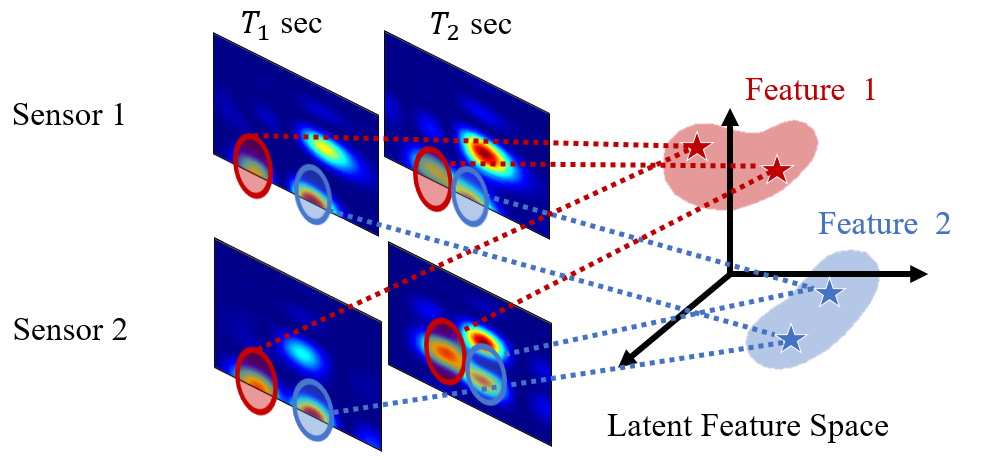}
    \caption{\emph{\textbf{Illustration of representation learning:}} For distinct, 
    nearby wireless channels, a \emph{feature representation} maps channels between two 
    sensors. For example, we extract environment specific information 
    (red circle marker on the upper-left spectrum) and map it as Feature~1. We then integrate
    this feature with Sensor~2's radio-specific information (red 
    circle marker on the bottom-left spectrum) to generate unseen channels.}
    \label{f:feature_embedding}
\end{figure}

\subsection{CLCP Model Design}
\label{s:design}


\begin{figure*}
\includegraphics[width=1\linewidth]{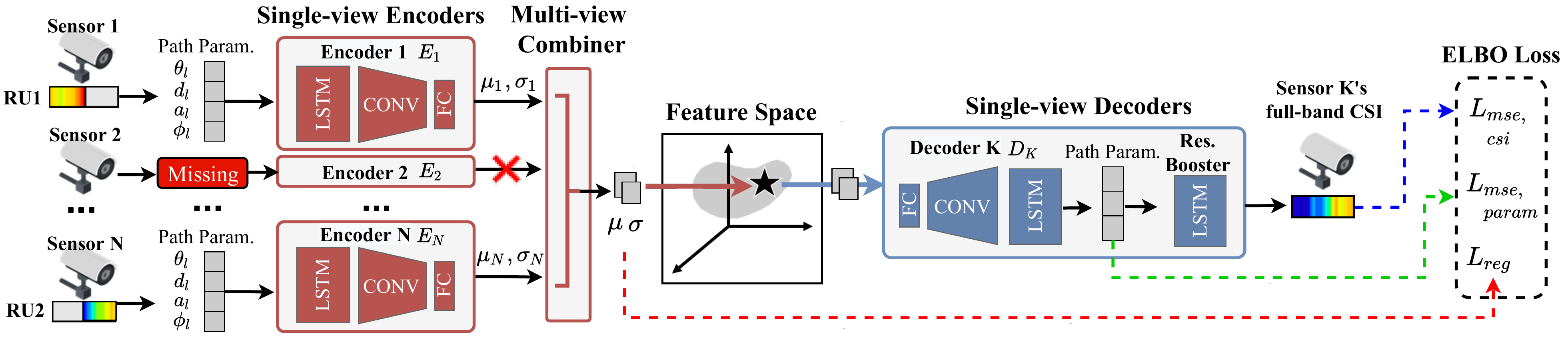}
\caption{\clcp{} ML model with $N$ measured channels, 
each represented as a set of wireless path parameters $\{\theta_{l}, d_{l}, a_{l}, \phi_{l}\}^{L}_{l=0}$ 
with $L$ paths estimated from measured channels. 
Each set of the parameters are served by 
a \textbf{Single-view Encoder} network $E_i$ ($i \in [1, N]$) 
that compresses the measured wireless path information of its dedicated radio and 
outputs variational parameters $\mu_{i}$ and $\sigma_{i}$. 
The \textbf{Multi-view Combiner} integrates 
all variational parameters into $\mu$ and $\sigma$, based on which 
\textbf{Single-view Decoder} 
networks $D_{K}$ generate a set of path parameters that are unobserved.
If any input channel is not observed, \shortname{} drops the respective encoder network ($E_{2}$, for example).}
\label{f:modules}
\end{figure*}

In the context of learning, 
cross-band channel prediction \cite{R2F2-sigcomm16, bakshi2019fast} is a
markedly different problem than cross-link channel prediction.
In the former, uplink and downlink channels 
share exactly the same paths in the wireless channel. 
Therefore, the learning task is simply to map the (complicated) effect of 
changing from one frequency band to another, given a fixed set of
wireless channel paths.
For cross-link channel prediction, on the other hand, 
the channels of nearby radios have distinct paths, and the 
learning task is to elucidate the correlations between the two links.
Since two different links share \textit{some} 
views on the wireless environment as shown in \cref{f:feature_embedding},
our learning task is to first discard radio-specific information 
from observed channels and extract features, 
representing information about link dynamics like moving reflectors. 
Then we integrate those extracted features with
radio-specific properties of the users, to synthesize unseen channels.
The first task is hard to accomplish because radio-specific 
and environment-specific information in the channel superpose 
in CSI readings and thus are not easily separable.
Therefore, we exploit representation learning to capture a
meaningful representation from CSI observations.  

Our \clcp{} ML model is summarized in \cref{f:modules}: 
there is a \emph{single-view encoder}
network $q_{\phi}(z|h)$ (\S\ref{s:sve}) dedicated to each single 
\emph{view} 
(\textit{i.e.}, channel $h$ of each radio). 
Every encoder outputs the distribution parameters, $\mu$ and $\sigma$, and a 
\emph{multi-view combiner} (\S\ref{s:mvc}) fuses all output parameters into a joint 
low-dimensional representation $z$. 
When channel is not observed, we drop its respective encoder network 
(\textit{e.g.} $E_{2}$ in \cref{f:modules}). 
A decoder network $p_{\theta} (h|z)$ (\S\ref{s:svd}) serves each single 
view of a \emph{target} radio whose CSI we seek to synthesize. 
Each decoder samples a data point from the joint latent representation $z$ to 
construct a \emph{cross-link predicted} channel. 

A key challenge in designing \clcp{} is 
that across different prediction instances, the channel inputs vary 
as we exploit channels in existing OFDMA transmission as an input.
In \cref{f:modules}, the input channels are two RUs from a previously acquired OFDMA packet, 
which were assigned to Sensor $1$ and Sensor $N$, respectively.
At the next prediction instance, 
an OFDMA packet is likely to contain a different set of RUs, each assigned to a different radio.
This inconsistency makes the learning process highly complex.
We will address how to make \clcp{} robust against the observations that vary with respect to frequency (\S\ref{s:param}) and link (\S\ref{s:mvc}).


\subsubsection{Path Parameter Estimator}
\label{s:param}
\clcp{} aims to infer geometric transformations 
between the physical paths traversed by different links. 
We reduce learning complexity of \clcp{} by extracting the geometric information (\textit{i.e.} wireless path parameters) from raw CSIs and directly using them as input data.
More importantly, the path parameters are \textit{frequency-independent}. 
Hence, using the path parameters
makes the model robust against the observation with varying combination of RUs.
Specifically, we represent channel $h$ observed at antenna $M_{i}$
as a defined number of paths $L$, 
each described by an arrival angle \emph{$\theta_{l}$}, 
a time delay \emph{$d_{l}$}, an attenuation \emph{$a_{l}$}, and a reflection \emph{$\phi_{l}$} as follow:
\begin{equation}
    h_{M_{i}, \lambda} = \sum_{l}^{L}(a_{l}e^{\frac{-j2\pi d_{l}}{\lambda}+j\phi_{l}})e^{\frac{-j2\pi ikcos(\theta_{l})}{\lambda}}
\label{eq:channel}
\end{equation}
where $\lambda$ and $k$ are wavelength and antenna distance. 
To extract the 4-tuple of parameters $\{(\theta_{l},d_{l},a_{l},\phi_{l})\}^{L}_{l=0}$, 
we use maximum likelihood estimation.
For simplicity, we now denote the 4-tuple as $\ddot{h}$.

\subsubsection{\shortnames{} Single-View Encoder}
\label{s:sve}
The goal of each \shortname{} single-view encoder $q_\phi(z|\ddot{h})$ is to learn an efficient 
compression of its corresponding view (\textit{i.e.} wireless path parameters) into a low-dimensional feature.
Like VAE, it encodes the corresponding channel into Gaussian distribution parameters, $\mu$ and $\sigma$, for better generalizability.
Each of single\hyp{}view encoder consists of 
the long short term memory layer (LSTM) 
followed by two-layer stacked convolutional layers (CNN) and 
fully connected (FC) layers. 
In each layer of CNNs, 1D kernels are used
as the filters, followed by a batch norm layer that normalizes
the mean and variance of the input at each layer. 
At last, we add a rectified linear unit (ReLU) to 
non\hyp{}linearly embed the input into the latent space. 
We learn all layer weights of the encoders and decoders end-to-end through backpropagation.
For the links that are not observed, we drop the respective encoder networks. 
For example, in \cref{f:modules}, Radio~2 was not a part of OFDMA transmission when \clcp{} initiates prediction.
Therefore, \shortname{} simply drops the single-view encoder dedicated to Radio~2.

\label{s:poe}

\subsubsection{\shortnames{} Multi-view Combiner}
\label{s:mvc}
A na\"{\i}ve approach to learn from varying multi-view inputs 
is to have an encoder network for each combination of input.
However, this approach would significantly increase the number of 
trainable parameters, making \shortname{} computationally intractable.
Using a multiview combiner, we assign one encoder network per each view and 
efficiently fuse the latent feature of all $N$ encoders into a joint representation. 
We model the multiview combiner after the \emph{product-of-experts} (PoE)
\cite{hinton2002training, wu2018multimodal} whose core idea is 
to combine several probability distributions (\emph{experts}), 
by multiplying their density functions, allowing 
each expert to make decisions based on a few dimensions 
instead of observing the full dimensionality.
PoE assumes that 
$N$ inputs are conditionally independent given the latent feature $z$, a valid
assumption since the channels from different radios are conditionally independent due 
to independently fading signal paths.
Let encoder networks
$q_{\phi}(z|\mathbf{\ddot{H}})$ for each subset of input channels 
$\mathbf{\ddot{H}} =\left \{  \right.\ddot{h}_{i}$ | channel of $i^{\mathrm{th}}$ radio $\left.  \right \}$. 
Then with any combination of the
measured channels, we can write the joint posterior distribution as:
\begin{equation}
 q_{\phi}(z|\mathbf{\ddot{H}})\propto p(z)\prod _{\ddot{h}_{n}\in \ddot{H}}\widetilde{q}(z|\ddot{h}_{n})
 \label{eq:poe}
\end{equation}
where $p(z)$ is a Gaussian prior, and $\widetilde{q}(z|\ddot{h}_{n})$ is an encoder 
network dedicated to the $n^{\mathrm{th}}$ radio.
\Cref{eq:poe} shows that we can approximate the distribution 
for the joint posterior as a product of individual posteriors.
The foregoing conditional independence assumption allows factorization of 
the variational model as follows:
\begin{equation}
    p_{\theta}(\ddot{h}_{1},\dots,\ddot{h}_{N},z) = p(z)p_{\theta}(\ddot{h}_{1}|z)p_{\theta}(\ddot{h}_{2}|z)\dots p_{\theta}(\ddot{h}_{N}|z). 
\end{equation}
where $p_{\theta}(\ddot{h}_{1},\dots,\ddot{h}_{N},z)$ is a \shortname{} model, and $\ddot{h}_{N}$ is the channel of Radio $N$.
With this factorization,  we can
simply ignore unobserved links,
which we will later discuss in \cref{s:obj}.
Finally, we sample from the joint distribution parameters $\mu$ and $\sigma$ to obtain 
the representations $z$ where $z=\mu+\sigma\odot\epsilon$ and $\epsilon\sim\mathcal{N}(0,\mathbf{I})$.

\subsubsection{\shortnames{} Single-View Decoder}
\label{s:svd}
Our single-view decoder $p_\theta(\ddot{h}|z)$ is another DNN, parameterized by 
$\theta$, whose input is the joint representation $z$.
The goal of each decoder is to synthesize an accurate channel estimate of its dedicated radio.
The decoder architecture is in the exact opposite order of the encoder architecture. 
It consists of two-layer stacked CNNs and FCs followed by an LSTM. 
In each CNN layer, 1D kernels are 
used as the filters with a batch norm layer and ReLU for 
the activation function.  
The LSTM layer predicts the path parameters of a target radio, 
which in turn we use to construct a full-band channel based on \cref{eq:channel}.
In practice, estimating path parameters is 
likely to cause a loss of information to some extent. 
To compensate such loss,
constructed channel is fed to an extra neural network layer,
a \emph{resolution booster}, to generate a final channel estimate $h_{pred}$.  


\subsection{\large Objective function and training algorithm}
\label{s:obj}
Recall the objective function of the VAE in \cref{eq:vae_term}.
With $\mathbf{\ddot{H}} =\left \{  \right.\ddot{h}_{i}$ | 
    channel of $i${th} radio$\left.  \right \}$, 
our ELBO is redefined as:
\begin{equation}
\begin{aligned}
        \mathrm{ELBO}
         = \mathbb{E}_{q_{\phi}(z|\mathbf{H})}\left[\log\, p_{\theta}(\ddot{h}|z)\right] 
            -\beta D_{\mathrm{KL}}(q_{\phi}(z|\mathbf{H})||p(z))
\end{aligned}
\end{equation}
where $\beta$ is a weight for balancing the coefficient in the ELBO. 
Like VAE, the second term 
represents the 
regularization loss ($Loss_{reg}$) that makes the approximate joint posterior 
distribution close to the prior $p(z)$, which is Gaussian.
For the reconstruction error, we compute a mean squared error between the predicted and ground-truth channels as follows:
    $Loss_{mse, csi} = \frac{1}{S}\sum_{s=0}^{S} ( 
    \left \|h_{s,gt}- h_{s,pred} \right \|_{2})$
where $S$ is the number of subcarriers and $h_{gt}$ is the ground-truth CSI.
Besides the CSI prediction loss, we compute the intermediate path parameter loss, 
which is a mean squared error between the predicted and ground-truth path parameters.
However, some paths are stronger than others when superimposed.
Hence, we weight the error of each path based on its path amplitude $a$ as follow:
$Loss_{mse, param} = \sum_{l=0}^{L} (a_{l}  
    \left \|\ddot{h}_{l,gt}- \ddot{h}_{l,pred} \right \|_{2})$. 
Then, the first term of our $\mathrm{ELBO}$ becomes $Loss_{mse}=-(\alpha Loss_{mse,csi}+\eta  Loss_{mse,param})$ where $\alpha$ and $\eta$ are weight terms.
Finally, negating $\mathrm{ELBO}$ defines
our loss function: $\boldsymbol{Loss}_{\mathrm{clcp}} = - \mathrm{ELBO}$.
By minimizing $\boldsymbol{Loss}_{\mathrm{clcp}}$, 
we are maximizing the lower bound of the probability of generating the accurate channels. 
\begin{figure}
    \includegraphics[width=.95\linewidth]{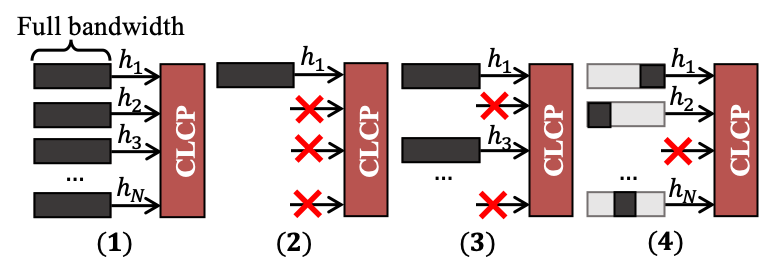}
    \caption{Multi-step training paradigm.}
    \label{f:training}
\end{figure}

\parahead{Multi-stage training paradigm.}
To adopt varying channel inputs, 
\shortname{} employs a special multi-step training paradigm~\cite{wu2018multimodal}.
If we train all encoder networks altogether,
then the model is incapable of generating accurate prediction 
when some links are unobserved at the test time.
On the other hand, if we individually train each encoder networks, 
we fail to capture the relationship across different links. 
Therefore, we train \clcp{} in multiple steps as shown in \cref{f:training}. 
First, our loss function consists of three ELBOs:
(1) one from feeding all $N$ full-band channel observation, 
(2) the sum of $N$ ELBO terms from feeding each full-band channel at a time, and
(3) the sum of $k$ ELBO terms from feeding $k$ randomly chosen subsets of full-band channel, $H_{k}$.
We then back\hyp{}propagate the sum of the three to train all networks end-to-end.
Lastly, (4) we repeat this 3-step training procedure 
with random subset of subcarriers to mimic channels in actual OFDMA transmission.

\begin{figure}
\centering
    \centering
    \includegraphics[width=.9\linewidth]{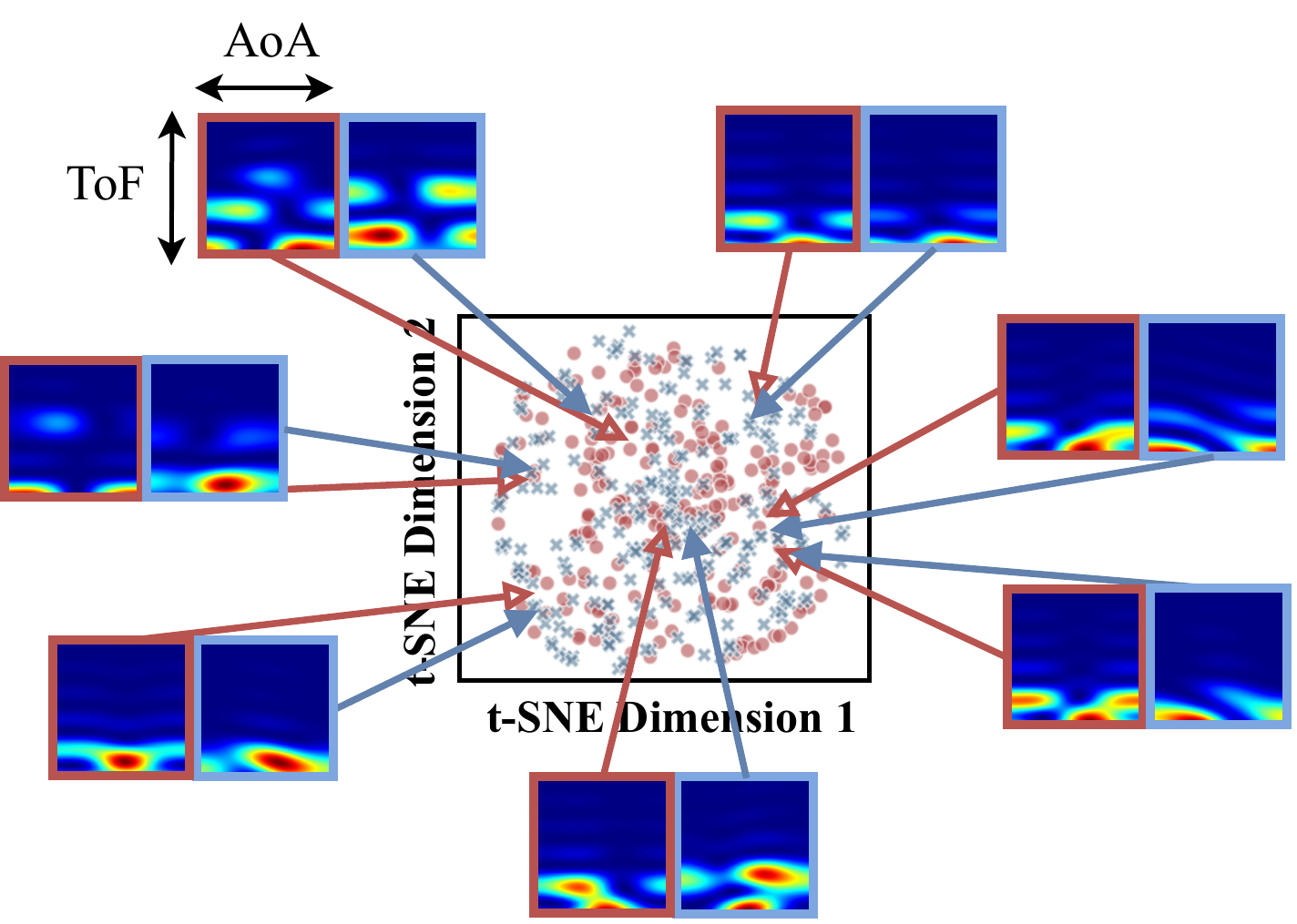}
    \caption{\textbf{\emph{\shortname{} explainability:}} A fully trained latent space example with 2D t-SNE visualization. The estimated path parameters, ToF for y-axis and AoA of x-axis. Encoded CSI instances of Radio 1 are highlighted in red with empty pointer and Radio 2 are colored in blue with a filled pointer.}
    \label{f:latent}
\end{figure}

\subsection{Cross-Link Joint Representation}
In the latent space, the closer two encoded low-dimensional features are to each other,
the more relevant they are for a given task.
Assume that we encoded the channels of two radios into our low-dimensional features. 
If these features are closely located in the latent space,
it is likely that the two nearby links have been affected by the 
same moving reflectors, simultaneously. 
By visualizing the low\hyp{}dimensional features in the latent space 
with the path parameters, 
we attempt to get some insight on the learning mechanism of \shortname{}.
For visibility, we reduce the dimensionality of the latent space into two
by performing t-SNE~\cite{maaten2008visualizing} dimension reduction technique.
Specifically, we collected the channel instances of two radios for $3$ hours and 
randomly selected these channels at $250$ different time-stamps. 
Then we fed the selected channels into the \shortname{} encoders.
Each encoder output $\mu$ is represented by a color\hyp{}coded data point where
red and blue data point denote a low-dimensional feature of Radio 1's channel and Radio 2's channel, respectively. 

\cref{f:latent} provides an in-depth analysis on the fully trained latent embedding via two wireless 
path parameters, Time-of-Flight and Angle-of-Arrival. 
For each radio, low-dimensional features 
are closely located when their corresponding path parameters are similar. 
For example, Radio 1's two low-dimensional features on right are in close proximity, 
and their corresponding path parameters resemble each other.
More importantly, we can also observe a pattern across different radios.
Although two radios have distinct path parameter values, 
the number of strong reflectors shown in Radio 1's spectrum and Radio 2's spectrum are similar when their low-dimensional features are close. 
For instance, the upper-left spectrum shows a lot of reflectors for both Radio 1 and 2 while the bottom-left spectrum has only one for both.
These observations demonstrate that the model is capable of properly encoding the wireless channels and
distributing the encoded features based on their relevance, which depends on the movement of reflectors. 
Also, when unseen channels are fed to the model, the model can still locate the encoded feature onto the latent space and make a good generalization based on the prior instances.

\subsection{Scheduling and Resource Allocation}
\label{s:sra}
\label{s:appendix}

\begin{figure}[t]
\centering
\includegraphics[width=1\linewidth]{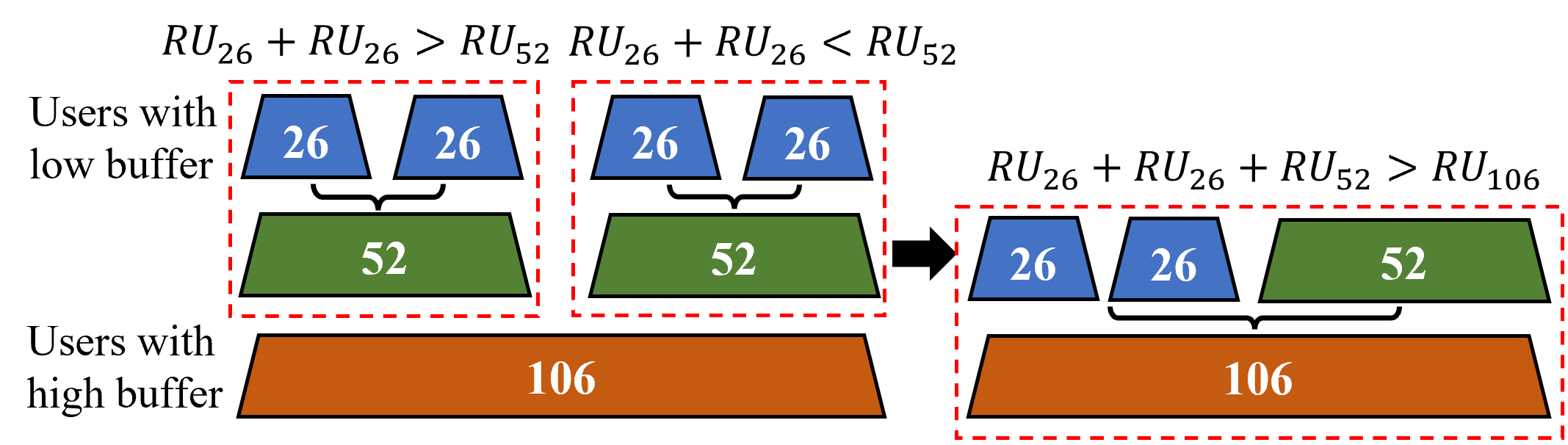}
\caption{Divide-and-conquer scheduling and resource alloction.}
\label{f:sra}
\end{figure}

\begin{figure*}[t]
\begin{subfigure}[b]{0.285\linewidth}
\includegraphics[width=\linewidth]{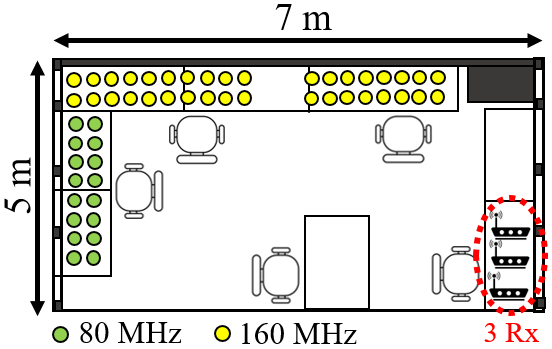}
\caption{Testbed for cashierless store.}
\label{f:testbed:testbed2}
\end{subfigure}
\begin{subfigure}[b]{0.44\linewidth}
\includegraphics[width=\linewidth]{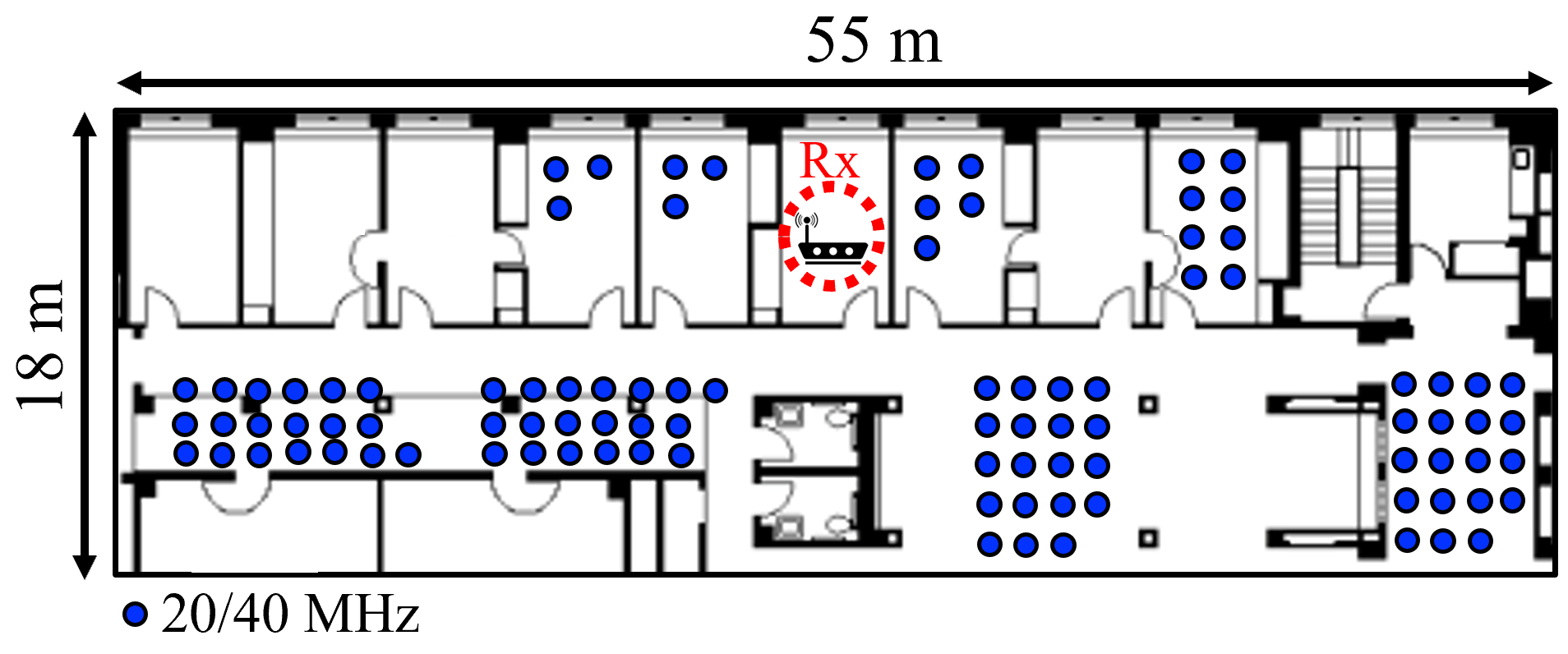}
\caption{Testbed for smart warehouse.}
\label{f:testbed:testbed1}
\end{subfigure}
\begin{subfigure}[b]{0.265\linewidth}
\includegraphics[width=\linewidth]{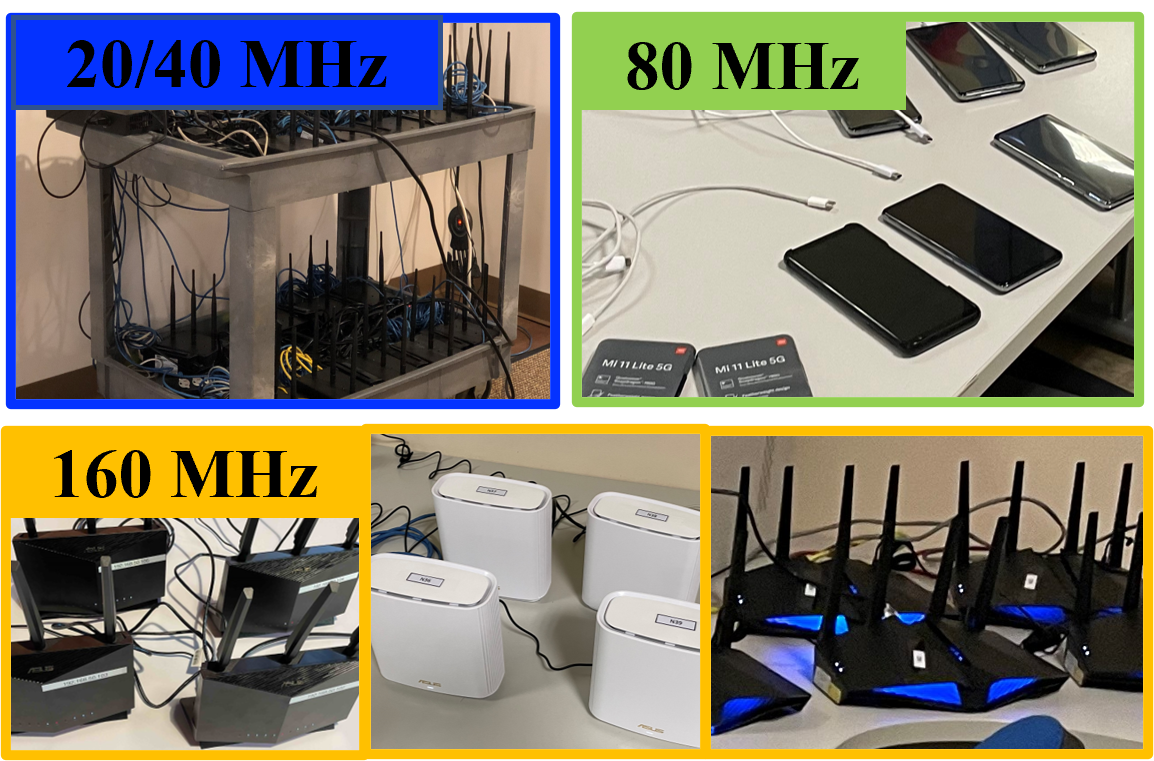}
\caption{Hardware devices.}
\label{f:testbed:hardware}
\end{subfigure}
\caption{\clcp{} preliminary experimental testbed floor plans 
and radio hardware with different operating bandwidths.}
\label{f:testbed}
\end{figure*}

Our scheduling algorithm exploits both channel conditions and buffer status to compute an optimal user schedule. 
In OFDMA, channel bandwidth is divided into RUs with various sizes from the smallest $26$ tones ($2$ MHz) up to $996$ tones ($77.8$~MHz). The size and the locations of the RUs are defined for $20$, $40$, $80$, and $160$~MHz channels. 
Our goal is to select an optimal set of RUs: one that covers 
the entire channel bandwidth, and maximizes the sum channel capacity. 
At the same time, we must consider the buffer status of all users, such that devices that require a lot of data, like streaming video, can be assigned a large RU, while devices that require very little data can be assigned a small RU. 
Scheduling is challenging as the size of the search space increases exponentially with the number of users and RU granularity.

To efficiently compute the optimal set of RUs, we reduce the search space by constraining the user assignment for each RU based on the user buffers and adopt a divide-and-conquer algorithm~\cite{wang2018scheduling} to quickly compute the optimal RU combination that maximizes channel capacity. 
Given the RU abstraction shown in \Cref{f:sra}, we first search for a user that maximizes the channel capacity for the first two $26$-tone RUs. Since this RU size is small, 
only one of the users with low buffer occupancy is chosen for each RU. 
Then, we select a best user for the $52$-tone RU from a group of users with 
moderate buffer occupancy and compare its channel capacity with the 
sum of two $26$-tone RUs capacities. 
This step repeats for the subsequent resource blocks, and the RU combination with higher capacity is selected and compared with larger RUs until the combination completes the full bandwidth. 

\section{Implementation}
\label{s:impl}

We conduct an experimental study on cross-link channel prediction in a large indoor lab (\cref{f:testbed:testbed2}) and in an entire floor (\cref{f:testbed:testbed1}) for the cashierless store and the smart warehouse scenario, respectively.
Typical cashierless stores consist of cameras and smartphones that demand large amounts of traffic; hence, in \cref{f:testbed:testbed1}, we collect channel traces from high-bandwidth 802.11ax commodity radios.
Specifically, the three receivers highlighted in red are Asus RT-AX86U APs supporting 802.11ax, 4x4 MIMO operation, and $160$-MHz bandwidth (i.e., $2048$ subcarriers per spatial stream) at $5$ GHz.
The transmitting nodes, shown in \cref{f:testbed:hardware}, include several Asus RT-AX82U and Asus ZenWifi XT8 routers (each one with four antennas), as well as some smartphones, like the Samsung A52S (single-antenna) and the Xiaomi Mi 11 5G (with two antennas each).
While the bandwidth of the 11ax Asus routers is $160$ MHz, 
the smartphones' radios can only handle up to $80$ MHz bandwidth.
In total, we identify $144$ separate links (here, we are counting each spatial stream as a separate link). 
To extract the CSI from commodity 11ax devices, we used the \href{https://ans.unibs.it/projects/ax-csi/}{AX-CSI extraction tool}~\cite{axcsi21}.

Since IoT devices in smart warehouses generally demand less data traffic,
we collect traces with $20$ and $40$ MHz bandwidth CSI for the scenario in \cref{f:testbed:testbed1}.
Both the AP and transmitting nodes are 11n WPJ558 with the Atheros QCA9558 chipset and three antennas.
Moreover, the nodes are placed in $95$ locations in NLoS settings, and we extract traces using \href{https://wands.sg/research/wifi/AtherosCSI/}{Atheros CSI Tool}.
All routers and phones together are generating traffic constantly using \texttt{iperf}.
For both testbeds, people moved at a constant walking speed of one to two meters per second.
Since commodity 11ax devices do not allow OFDMA scheduling on the user side,
we run a trace-driven simulation using a software-defined simulator.
We implement \clcp{} using \href{https://pytorch.org/}{Pytorch}, and the model parameters include batch size of $16$ and learning rate of $5e^{-6}$. 
We employ Adam for adaptive learning rate optimization algorithm. 


\textbf{Channel measurement error.}
The presence of noise in the dataset may significantly affect convergence while training.
Hence, we want to minimize some notable channel errors. 
First, the packet boundary delay occurs during OFDM symbol boundary estimation.
Assuming this time shift follows a Gaussian distribution~\cite{speth1999optimum, xie2018precise}, averaging the phases of multiple CSIs within the channel coherence time compensates for this error.
Thus, the AP transmits three sequential pilot signals, and upon reception, the clients average the CSI phase of these signals and report an error-compensated CSI back to the AP. 
Second, to compensate for the amplitude offset due to the power control uncertainty error, we leverage the Received Signal Strength Indicator (RSSI), reported alongside CSI in the feedback. We detect the RSSI outliers 
over a sequence of packets and discard the associated packet. 
Then, we average the amplitude of the channel estimates.  
Lastly, non-synchronized local oscillators cause some carrier frequency offset. 
To minimize this error, we subtract the phase constant of the first receiving antenna across all receiving antennas. 
Since phase constant subtraction does not alter a relative phase change across antennas and subcarriers, we preserve the signal path information in CSI.


\section{Evaluation}
\label{s:eval}
We begin by presenting the methodology for our experimental evaluation
(\S\ref{s:methodology}), followed by 
the end\hyp{}to\hyp{}end performance comparison on throughput and power consumption (\S\ref{s:e2e_perf}).
Lastly, we present a microbenchmark on prediction accuracy,
channel capacity, packet error rate, and PHY\hyp{}layer bit rates (\S\ref{s:microbenchmarks}).

\subsection{Experimental methodology}
\label{s:methodology}
\parahead{Use cases.}
We evaluate \shortname{} in two use case scenarios: a cashierless store and a smart warehouse. 
\textbf{Cashierless stores} typically experience a high demand of data traffic 
as densely deployed video cameras continuously stream their data to the AP for product and customer monitoring.
To reflect a realistic cashierless store application, 
we configure all users to continuously deliver standard quality video of $1080$p using UDP protocol for trace-driven simulation.
Also, we leverage $80$ and $160$ MHz bandwidth for every uplink OFDMA packet.
In \textbf{smart warehouses}, IoT devices transmit relatively little data traffic and are widely and sparsely deployed compared to the cashierless store use cases. 
Hence, each uplink packet has $20$ and $40$ MHz bandwidth, and the users transmit UDP data in NLoS settings.

\parahead{Evaluation metrics.}
To quantify the network performance, we define uplink throughput as 
a number of total data bits delivered to the AP divided by the duration. 
We measure it for every $500$ ms. 
Moreover, to evaluate the power consumption, 
we report a total number of Target Wake Time (TWT) packets. 
By definition, TWT is 11ax's power-saving mechanism where the client devices sleep between AP beacons,
waking up only when they need to transmit the signal (\textit{e.g.,} uplink data transmission and channel report). 
When a client is not scheduled for uplink transmission nor reporting CSIs to AP, 
the AP does not trigger TWT packet for the corresponding client.
By doing so, it effectively increases device sleep time and helps to conserve the battery of IoT devices.


\parahead{Baselines.}
Our baselines follow 
sounding protocols 
in which the AP periodically requests BSRs and CSIs from all users. 
Upon receiving the NDP from the AP, all users calculate the feedback matrix for each OFDM subcarrier as follows~\cite{8672643}:
\begin{equation}
    \frac{\mathrm{CSI\;tones} \times \mathrm{CSI\;bits} \times \mathrm{Tx Antenna} \times 
    \mathrm{Rx Antenna} \times T_{c}}{\mathrm{Subcarrier\;Group} \times \mathrm{Feedback\;Period}}
\end{equation}
where $T_{c}$ signifies the wireless channel coherence time.
We use $8$-bit CSI quantization, a channel coherence time of $15$ ms, and subcarrier grouping of $4$.
The other control protocols we consider are BSR report ($32$ bytes), BSR poll ($21$ bytes), CSI poll ($21$ bytes), MU-RTS ($20$ bytes), CTS ($14$ bytes), TF ($28+(5 \times K)$ bytes), and BlockAck\fshyp{}BA ($22+(5 \times K)$ bytes), where K denotes the number of users. Lastly, SIFS takes $10 us$. We note that BSRs and CSIs are delivered to the AP via OFDMA transmission to minimize the overhead.
\begin{figure}
        \centering
    \begin{subfigure}[b]{.72\linewidth}
    \includegraphics[width=\linewidth]{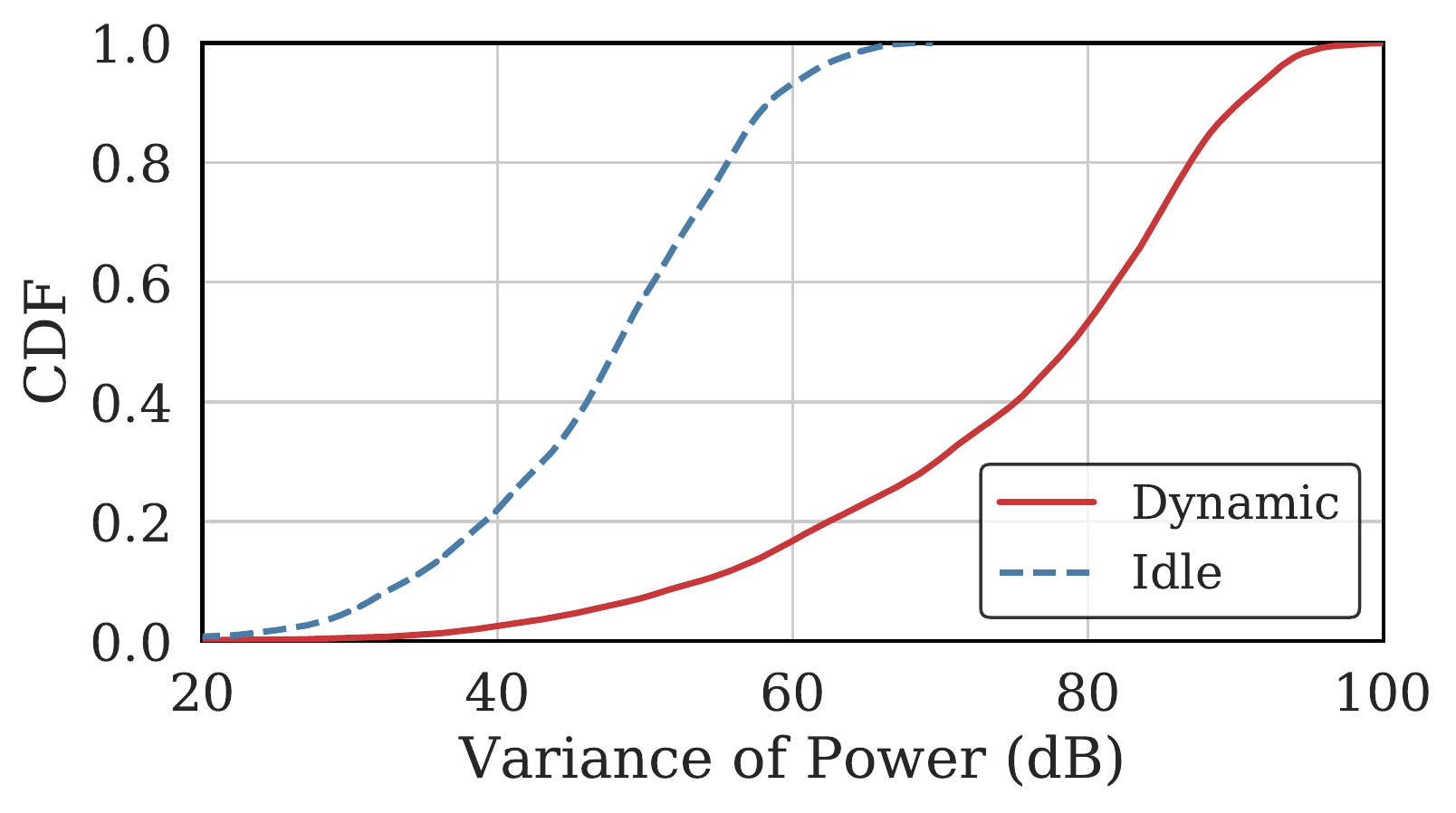}
    \end{subfigure}
    \caption{Channel variability. We indicate our channel data as a red line and idle channels as a dotted blue line.}
    \label{f:variance}
\end{figure}
\begin{figure*}[t!]
\begin{subfigure}[b]{0.375\linewidth}
\includegraphics[width=\linewidth]{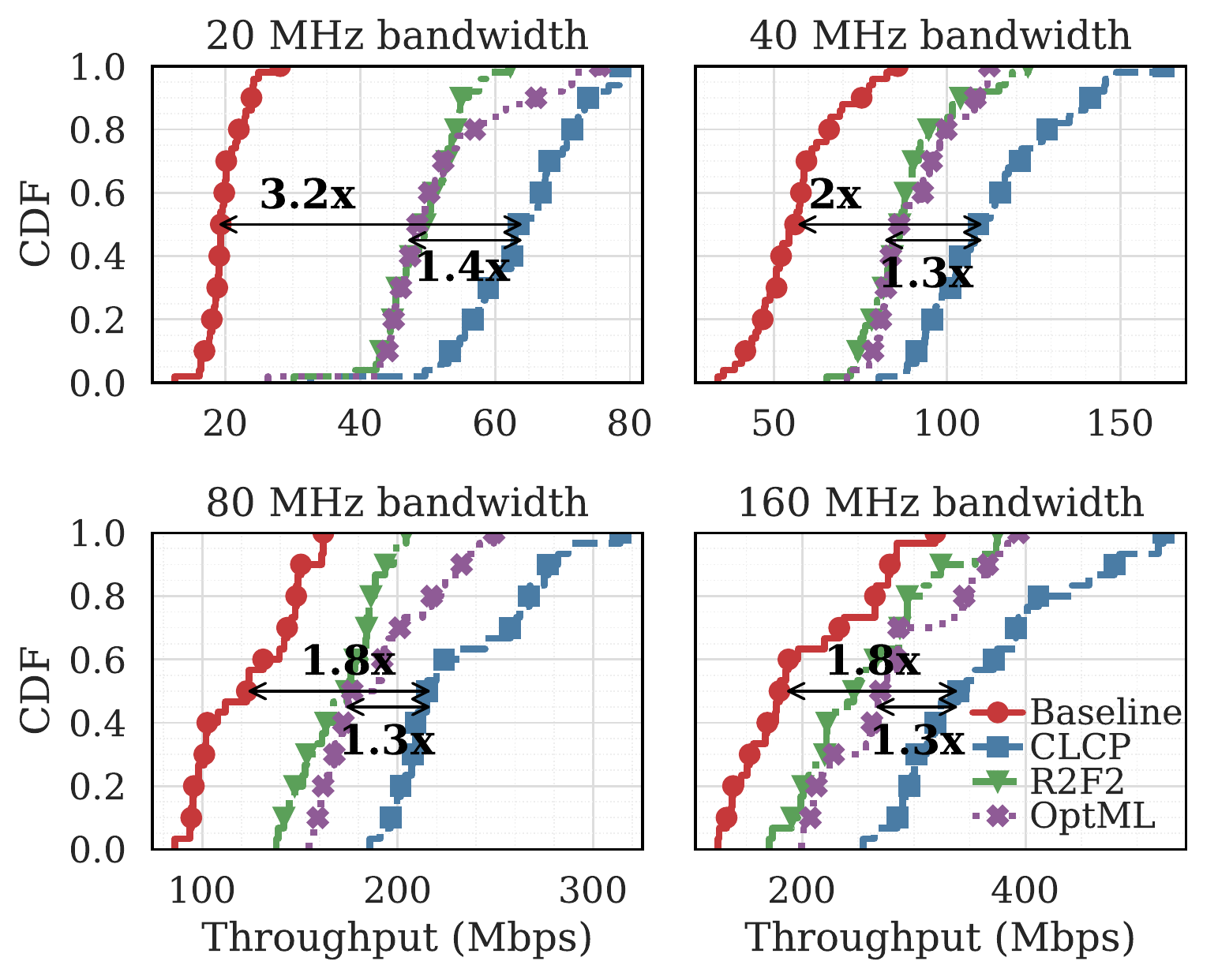}
\caption{Throughput performance across time.}
\label{f:eval:throughput_time}
\end{subfigure}
\begin{subfigure}[b]{0.375\linewidth}
\includegraphics[width=\linewidth]{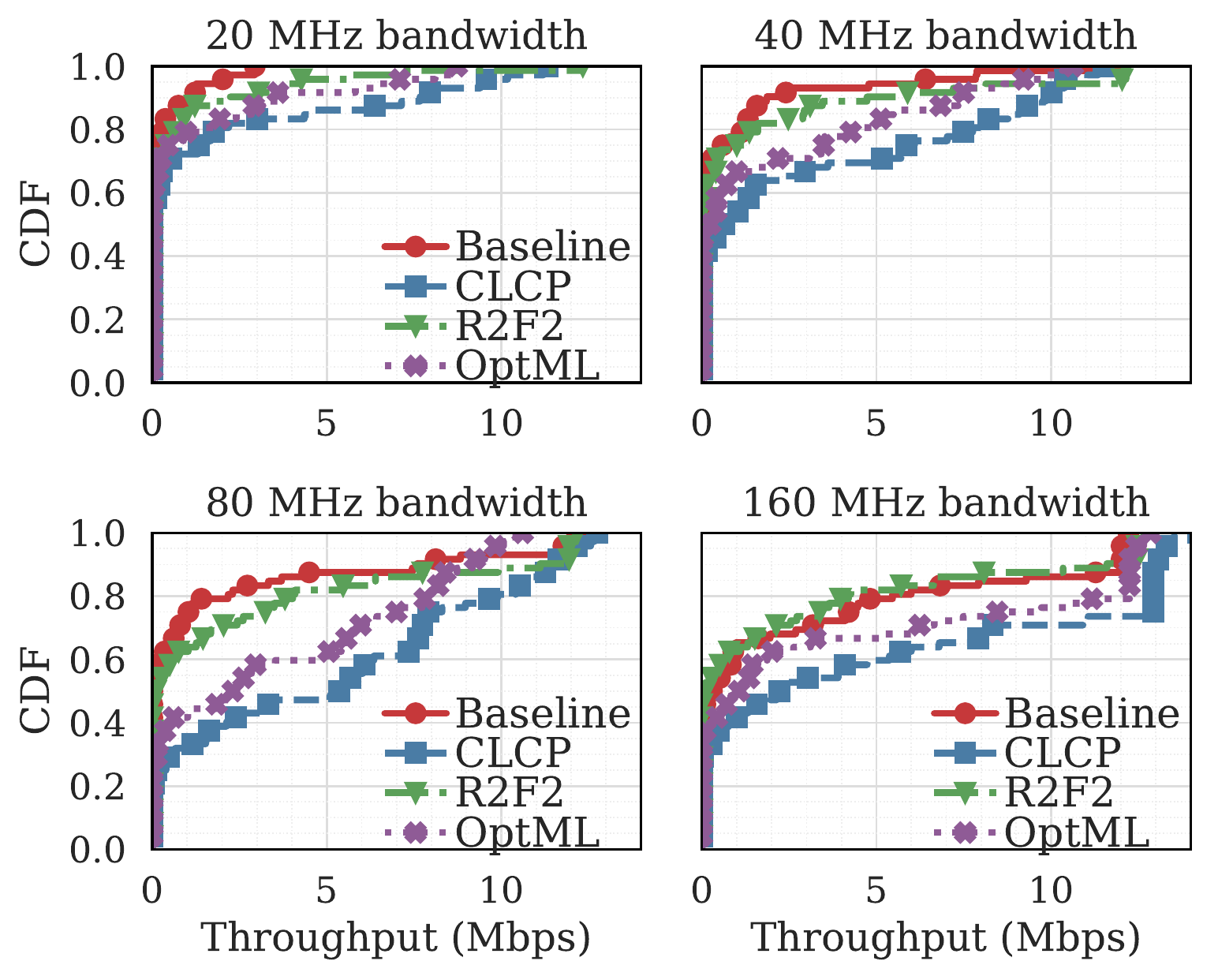}
\caption{Throughput performance across users.}
\label{f:eval:throughput_user}
\end{subfigure}
\begin{subfigure}[b]{0.2\linewidth}
\includegraphics[width=\linewidth]{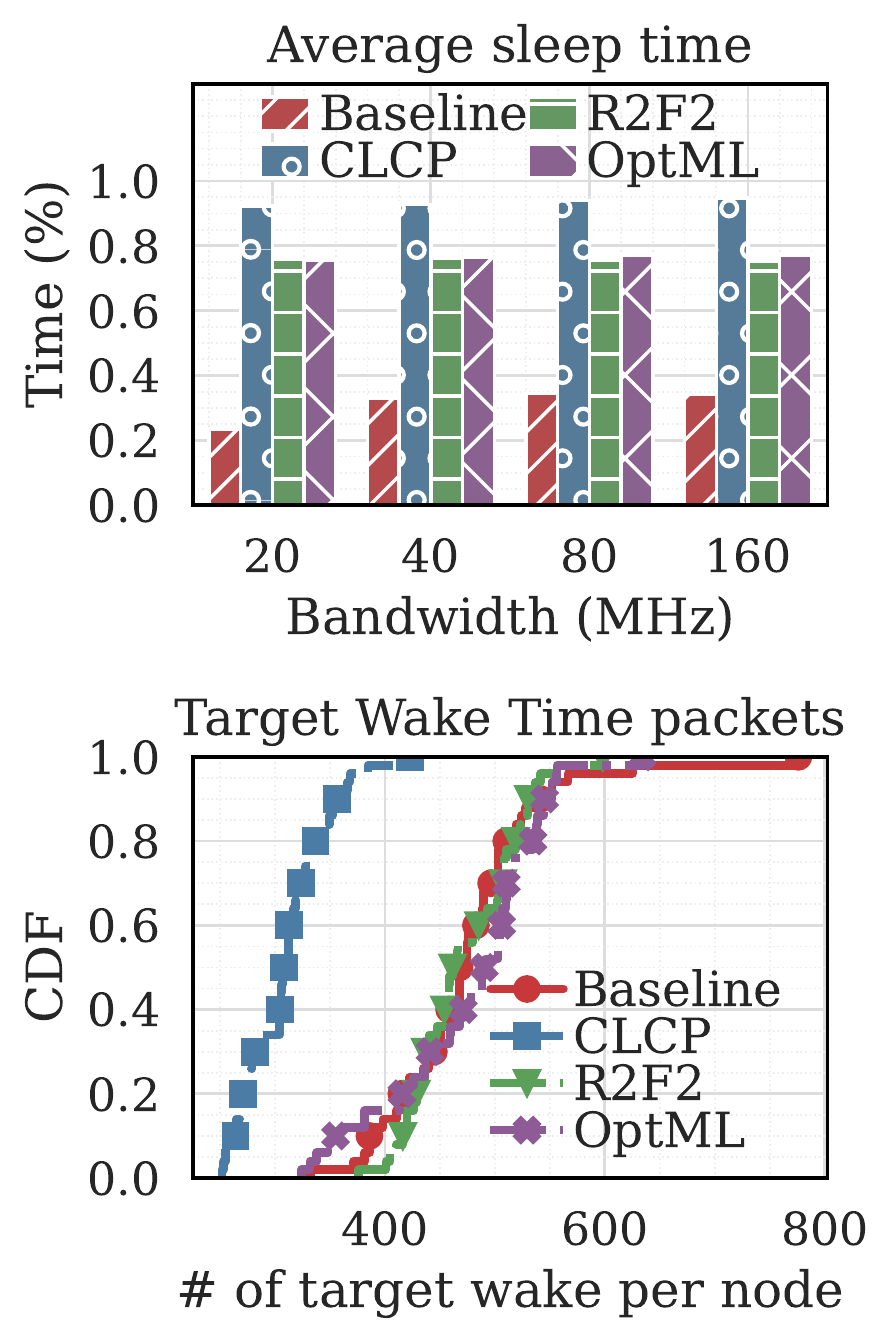}
\caption{Power consumption.}
\label{f:eval:power}
\end{subfigure}
\caption{End-to-end performance on throughput and power consumption: (a) aggregated throughput across time for every $500$ ms, (b) throughput across users for $20$, $40$, $80$, and $160$ MHz bandwidth, and (c) device sleep time over the entire transmission duration and the total number of Target Wake Time (TWT) triggered on every user.}
\end{figure*}

\parahead{Algorithms.} We compare \clcp{} to 
the following algorithms 
which collectively represent the state-of-the-art in channel prediction:
\begin{enumerate}[label=\arabic*)]
\item 
R2F2~\cite{R2F2-sigcomm16} infers downlink CSI for a certain 
LTE mobile\hyp{}base station link based on the path
parameters of uplink CSI readings for that \emph{same} link, 
using an optimization method in lieu of an ML\hyp{}based approach. 
\item OptML~\cite{bakshi2019fast} leverages a neural network 
to estimate path parameters, which, in turn, are used to 
predict across frequency bands. 
\end{enumerate}
Both algorithms predict downlink based on uplink channels in LTE scenarios, where
the frequencies for downlink and uplink are different.
To adopt these algorithms into our OFDMA scheduling problem instead, 
we use them to predict a full-band channel 
based on the RU in a received OFDMA packet.
We use a maximum likelihood approach for fast path parameter estimation.
For example, for $160$ MHz bandwidth, 
the AP triggers all clients to simultaneously transmit pilot signals in their $242$-subcarrier RUs.
Then the AP predicts the $2048$ subcarriers of the full band channel based on the received RUs.

For \shortname{}, we group IoT devices based on their proximity (3 to 5 m) and create one 
\shortname{} prediction module per group.
This is because wireless links that are far apart or separated by a wall have an extremely low correlation~\cite{charvat2010through}. 
However, since \shortname{} uses the latest OFDMA packet to make predictions, 
some groups might not have any of its users assigned to that OFDMA packet
and hence it is not possible to make predictions.
For these groups, we trigger uplink OFDMA packets and run cross-bandwidth channel prediction like R2F2 and OptML.

\parahead{Channel variability.}
We present 
the inherent variability of our channel environment.
\cref{f:variance} demonstrates the
variability of idle channels without human mobility in a dotted blue line and 
that of our channel environment affected by multiple moving reflectors in a red line.
Both channel environments are measured from all users in NLoS settings.
Precisely, we collect a series of channel readings over time and segment readings by one second duration.
For every segment and subcarrier of channels, we measure a power variance of channels over one second.
Then we generate the corresponding variance distribution, conveying the variances of all segments and subcarriers for each link
and average the distributions across all links. 
From \cref{f:variance}, we observe that power variance of our channel data is $\sim30$ dB higher than that of idle channel data.
This indicates that our links are not idle, and there is environment variability due to moving reflectors.


\subsection{End-to-end performance}
\label{s:e2e_perf}
In this section, we evaluate the end-to-end throughput performance of \shortname{} in comparison with the baseline, R2F2, and OptML across time and user. Then we demonstrate its performance on the power consumption.

\parahead{Significant throughput improvement.}
\Cref{f:eval:throughput_time} summarizes the end-to-end throughput performance of \clcp{} under $20$, $40$, $80$ $160$ MHz bandwidth channels. Each data of the curves indicates an aggregated uplink throughput within $500$ ms duration. 
With $20$ MHz bandwidth, \shortname{} improves the throughput by a factor of $3.2$ compared to the baseline and by a factor of $1.4$ for R2F2 and OptML. 
Similarly, \shortname{} provides $1.9x$ to $2x$ throughput improvement over the baseline for $40$, $80$, and $160$ MHz channels along with $1.3x$ improvement over R2F2 and OptML.  
Throughput improvements under $20$ MHz bandwidths are significant compared to larger bandwidths because delivering channel feedbacks overwhelm the network with a massive number of users and a small bandwidth. 
Hence, by eliminating the need for exchanging channel feedbacks, 
\shortname{} significantly improves spectral efficiency for smaller bandwidths.
Moreover, a maximum number of users allowed in $20$ MHz OFDMA is $9$ while $40$ MHz, $80$ MHz, and $160$ MHz allows $18$, $37$, and $74$ users for each OFDMA packet, respectively.
Hence, using OFDMA for delivering channel feedbacks with a small bandwidth is not as effective as sending them with a large bandwidth. 
Even with larger bandwidths, \shortname{} outperforms the baseline. 
Moreover, \shortname{} provides better throughput performance than two cross-band prediction algorithms, R2F2 and OptML.
While existing cross-band prediction algorithms require a pilot signal dedicated for channel sounding from all users, 
\shortname{} exploits the channel estimates obtained from existing transmissions and thus completely eliminates the need for extra signal transmissions and corresponding control beacons.

In \Cref{f:eval:throughput_user}, we present the end-to-end throughput performance across users. Here, each data indicates a throughput of one user within $10$ second duration of uplink traffics. 
It is worth noting that as the bandwidth increases, more users have an opportunity to send their data. 
Specifically, for $20$ MHz, only $20\%$ to $40\%$ of users send the data while for $160$ MHz, more than $50\%$ to $70\%$ of users communicate with the AP.  
More importantly, we observe that for all bandwidths, 
\shortname{} enables $15\%$ to $20\%$ more users to delivery their data within $10$ second duration by eliminating the channel sounding and increasing the spectral efficiency. 



\begin{figure*}[t!]
\begin{subfigure}[b]{0.24\linewidth}
\includegraphics[width=\linewidth]{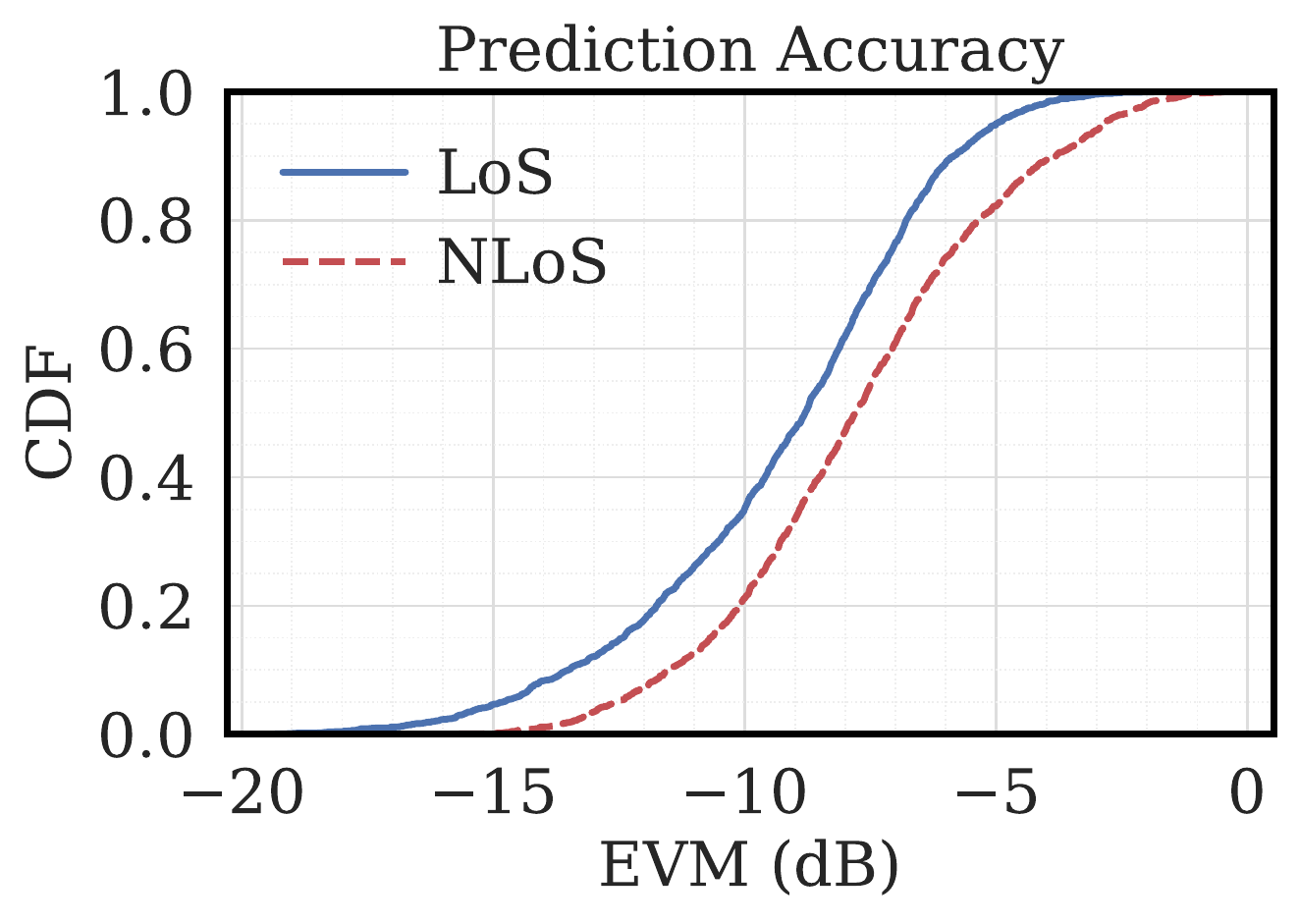}
\caption{Prediction accuracy.}
\label{f:accuracy}
\end{subfigure}
\begin{subfigure}[b]{0.245\linewidth}
\includegraphics[width=\linewidth]{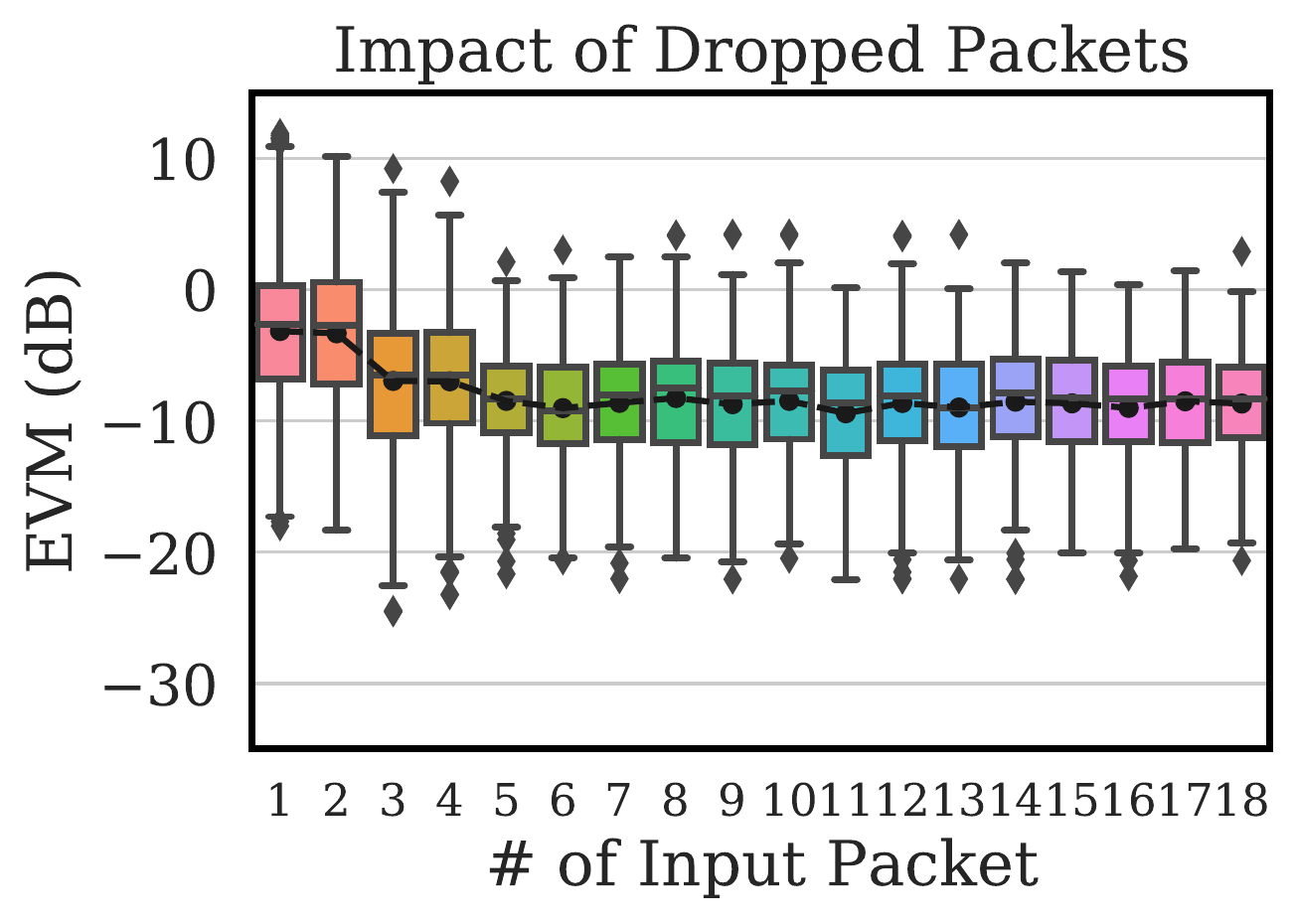}
\caption{Impact of input number}
\label{f:missing}
\end{subfigure}
\begin{subfigure}[b]{0.20\linewidth}
\includegraphics[width=1\linewidth]{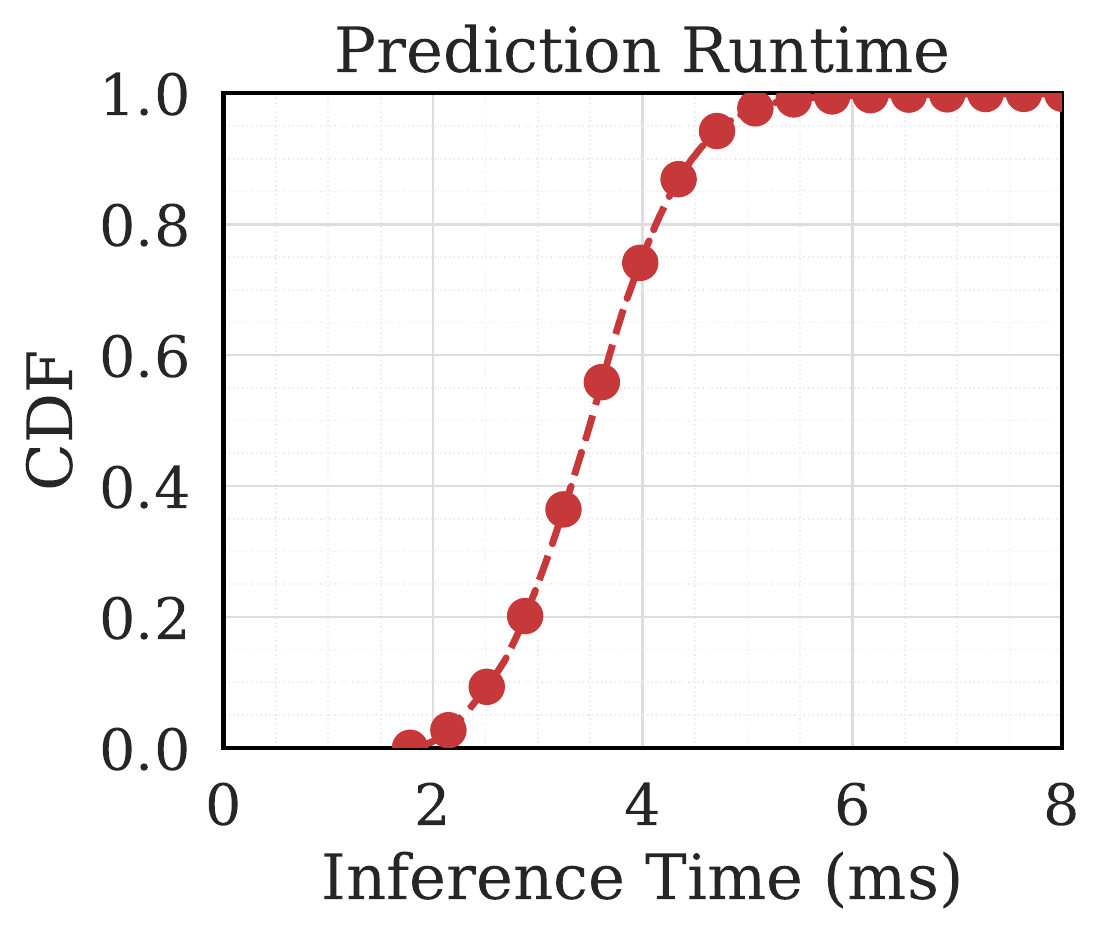}
\caption{CLCP runtime}
\label{f:runtime}
\end{subfigure}
\begin{subfigure}[b]{0.3\linewidth}
\includegraphics[width=\linewidth]{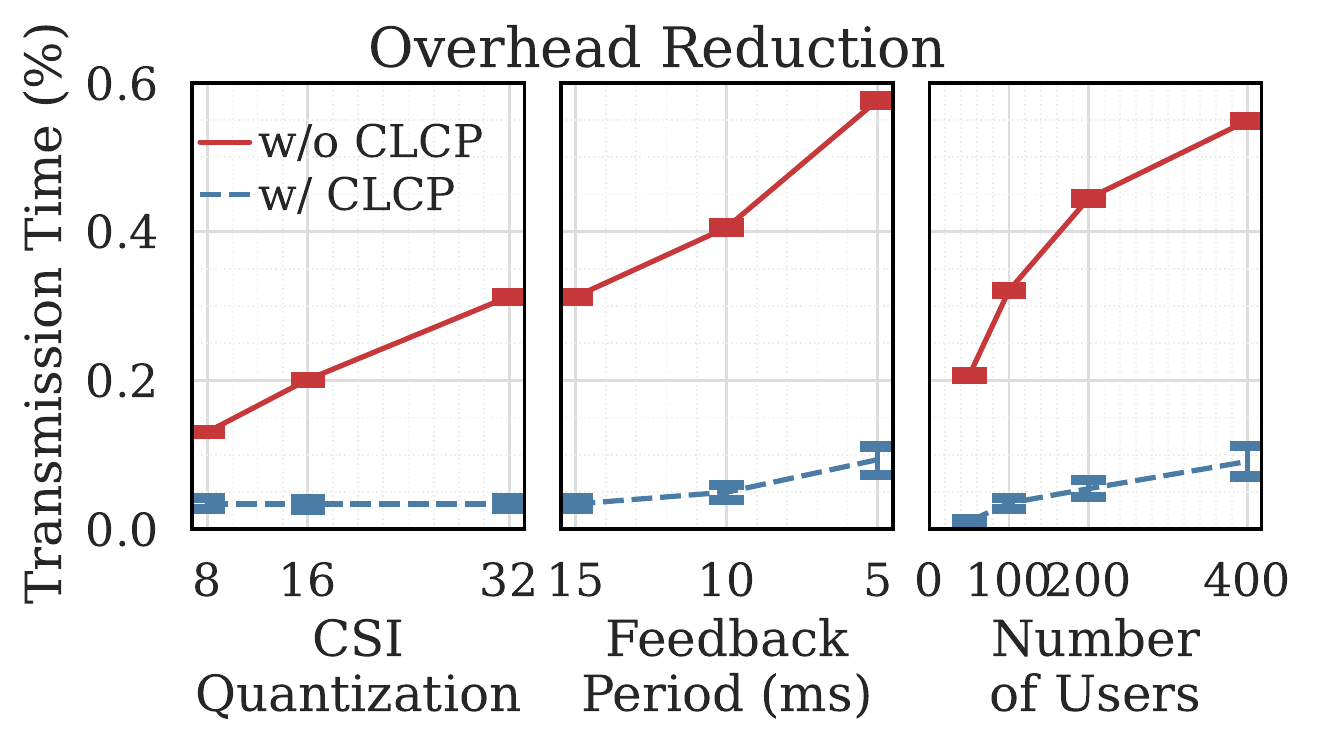}
\caption{Overhead reduction}
\label{f:overhead}
\end{subfigure}
\begin{subfigure}[b]{0.24\linewidth}
\includegraphics[width=\linewidth]{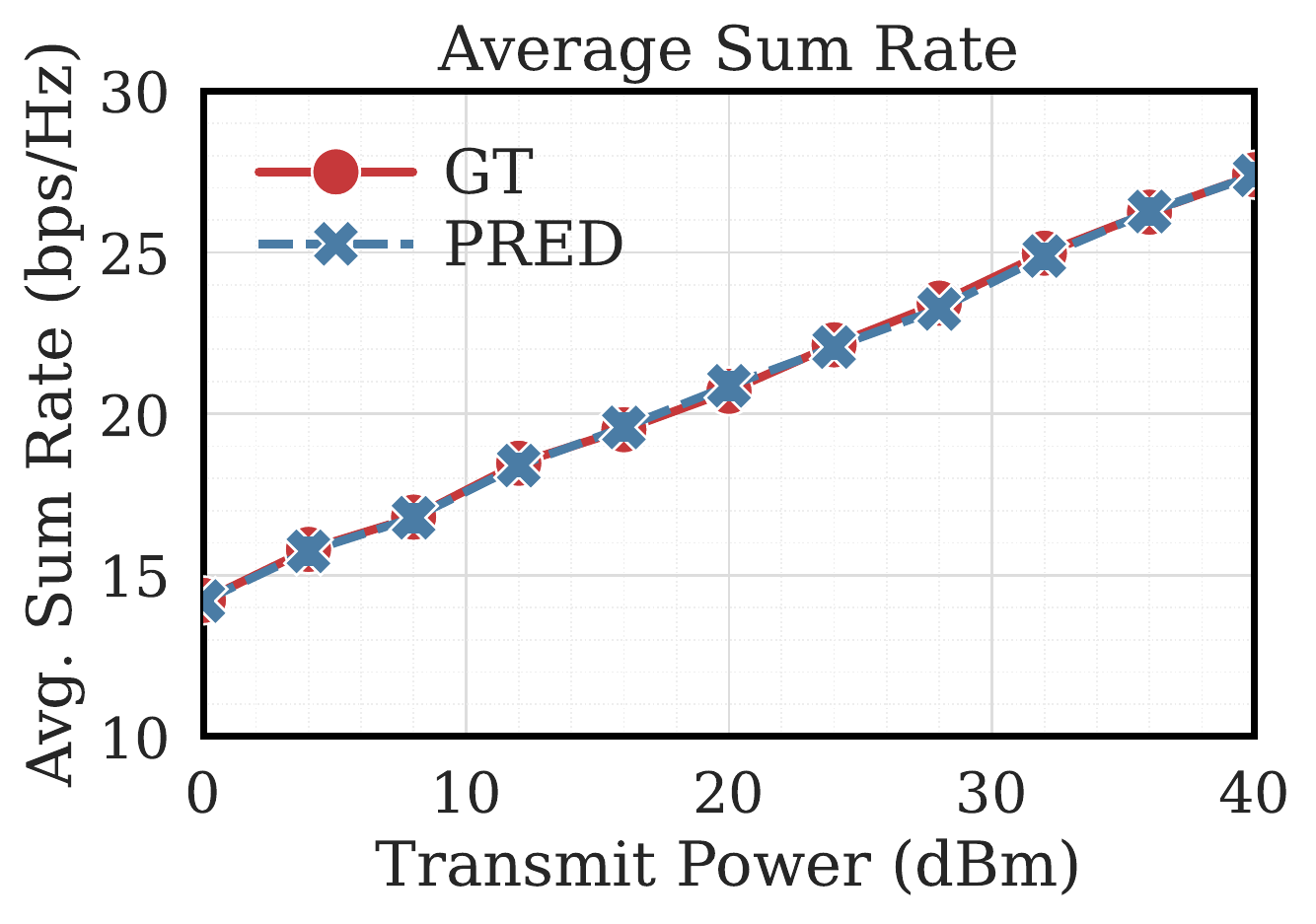}
\caption{Channel capacity.}
\label{f:sum_rate}
\end{subfigure}
\begin{subfigure}[b]{0.24\linewidth}
\includegraphics[width=\linewidth]{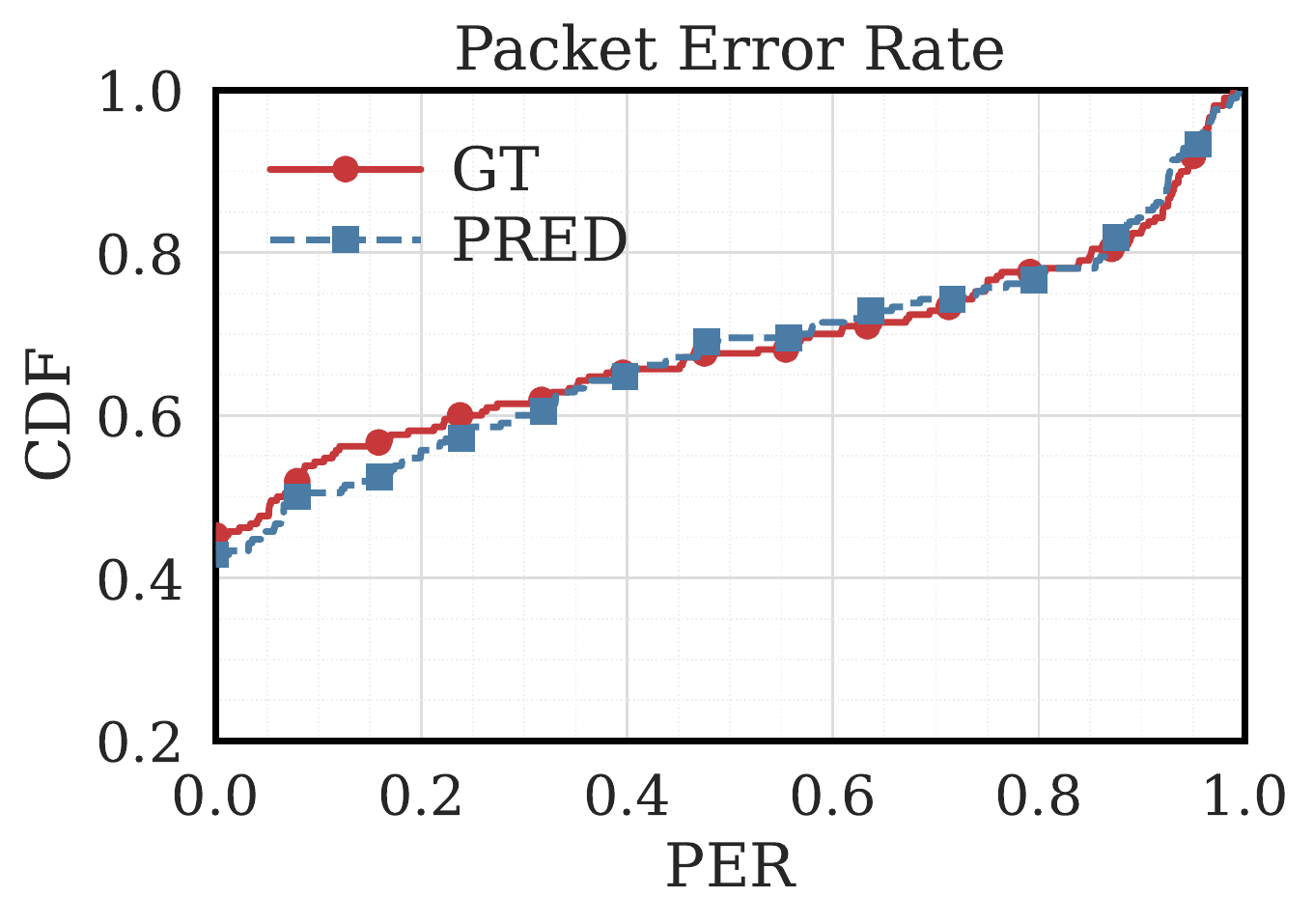}
\caption{PER distribution.}
\label{f:per}
\end{subfigure}
\begin{subfigure}[b]{0.24\linewidth}
\includegraphics[width=1\linewidth]{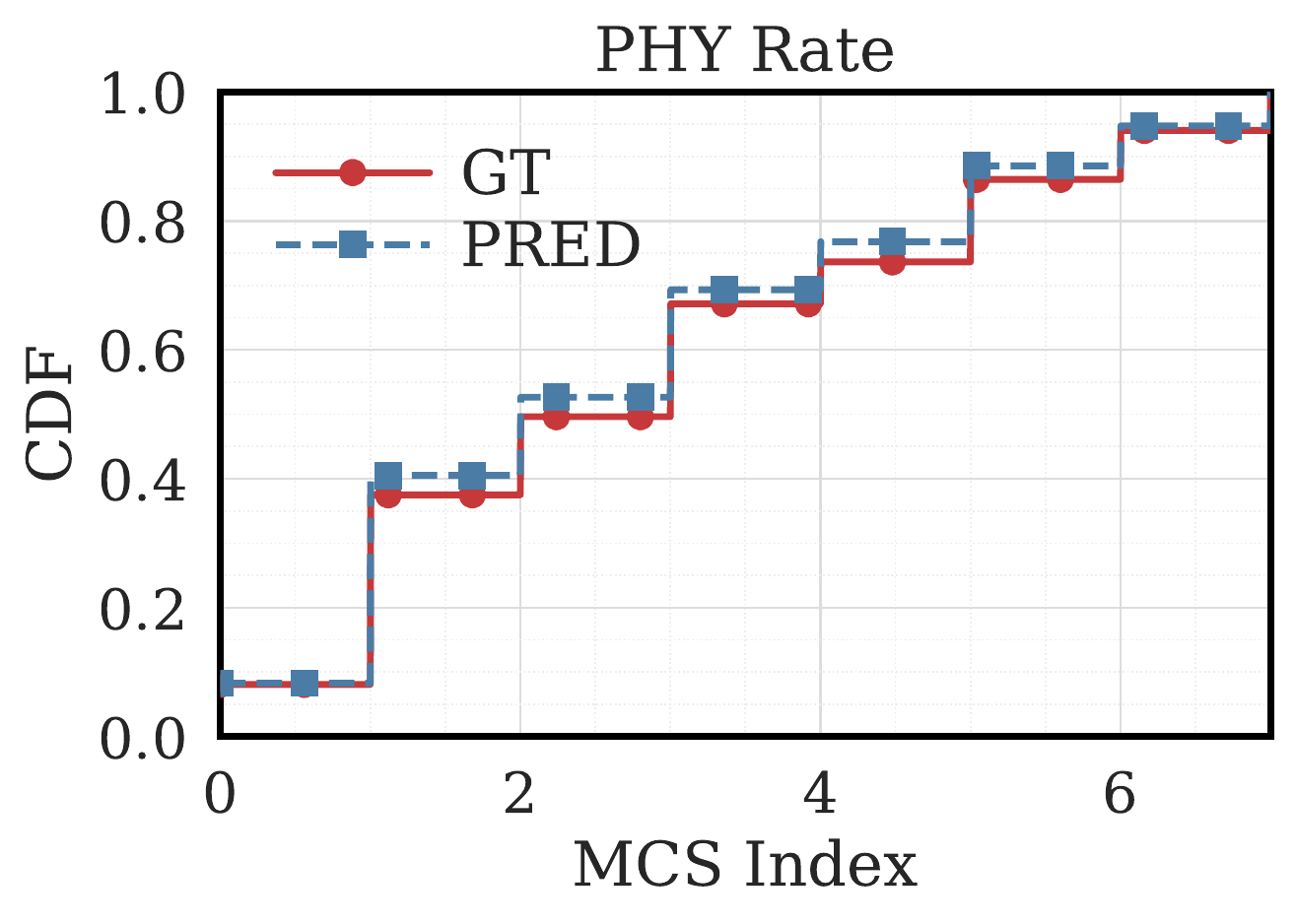}
\caption{PHY rate distribution.}
\label{f:phyrate}
\end{subfigure}
\begin{subfigure}[b]{0.25\linewidth}
\includegraphics[width=1\linewidth]{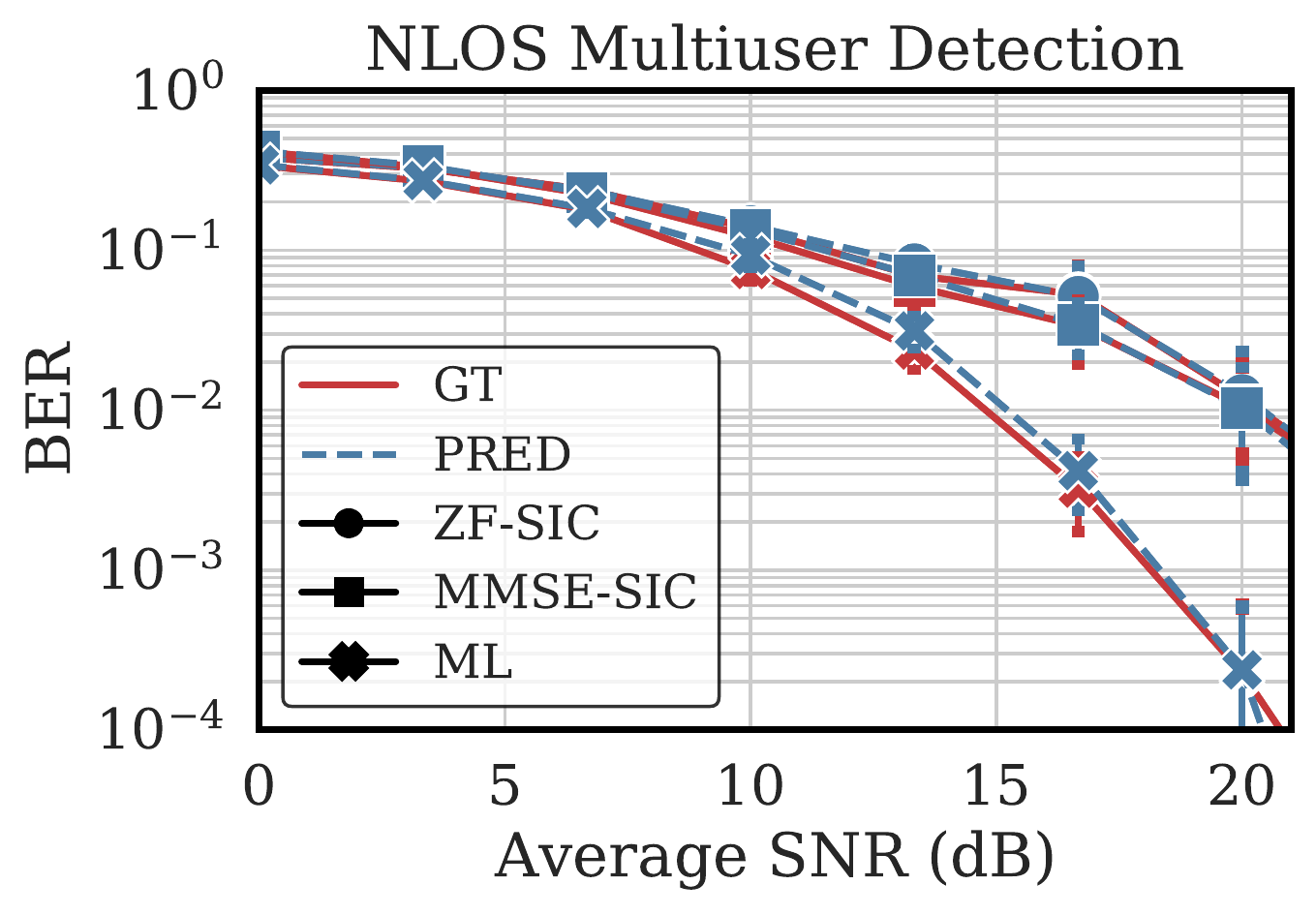}
\caption{BER distribution.}
\label{f:ber}
\end{subfigure}
\caption{Microbenchmark on the prediction accuracy, overhead reduction, and scheduling performance. 
}
\end{figure*}
\parahead{Increasing device sleep time.}
Target wake time (TWT) reduces the power consumption by letting users sleep and 
only waking up when they need to transmit their packets.
Thereby, users skip multiple beacons without disassociating from the network.
The subfigure on top of \Cref{f:eval:power} shows the average sleep time of all users over the entire measurement duration.
While the users sleep for slightly over $25\%$ of the time using the baseline algorithm, \shortname{} enables $90\%$ of users to remain in sleep, which is roughly $65\%$ and $15\%$ longer than the baseline and cross-band prediction algorithms, respectively.
We note that the average sleep time of the cross-band algorithms are much longer than the baseline.
This is because each user has to send at least $260$ byte of its channel information for the baseline 
while R2F2 and OptML simply transmit a pilot signal with some control overheads.
However, \shortname{} does not need bulky channel feedbacks nor pilot signals since the AP directly infers CSIs from existing transmissions.
This also helps minimizing contention between users and reduce the amount of time a user in power save mode to be awake.
The subfigure on bottom of \cref{f:eval:power} shows how many TWT packets are received by each user when $150$ MB data are delivered to the AP.
This is equivalent to how frequent each user is waking up from the sleep mode to participate in channel sounding and data transmission. 
Here, \shortname{} has significantly less TWT counts because the users stay idle during channel acquisition
while the baseline, R2F2, and OptML wake all users and make them to transmit a signal.
Given that the average wake power is $600$ uW and transmit power is $135$ mW per device, we can infer that the power consumption of \shortname{} is significantly smaller than the power consumption of the baseline, R2F2, and OptML. 

\subsection{Microbenchmark}
\label{s:microbenchmarks}
The first microbenchmark focuses on \shortname{}'s prediction performance 
across different users and varying number of observed channels.
We then analyze overhead reduction with varying parameters,
such as CSI quantization, feedback period, and number of users.
Lastly, we evaluate the performance of 
rate selection and multiuser detection using predicted CSIs.
Microbenchmark results are obtained under settings in \cref{f:testbed:testbed1}. 

\subsubsection{Prediction Accuracy}
As a measure of prediction accuracy, we use \emph{error vector magnitude} (EVM), 
which represents how far a predicted channel $H$ deviates
from a ground truth channel $H_{\mathrm{gt}}$:
$\mathrm{EVM} = 10 \log_{10}\left(|H - H_{\mathrm{gt}}|^2 / |H_{\mathrm{gt}}|^2\right)$.  
According to IEEE 802.11 specification~\cite{8672643, ieee2010ieee}, BPSK modulation scheme requires an EVM between $-5$ to $-10$ dB, and QPSK needs the EVM from $-10$ to $-13$ dB.  
In \cref{f:accuracy}, \shortname{} provides an average EVM of approximately $-8$ dB. 
Compared to the LoS setting, the NLoS setting shows a larger variation of EVMs across different users. This is because many wireless links in the NLoS setting are weak due to multiple wall blockages and long signal propagation distance. Such weak signals have low signal-to-noise ratio, and therefore the effect of noise is high and causing a high variation in prediction accuracy.

\noindent
\textbf{Impact of the number of observed channels.}
We evaluate prediction accuracy with varying number of observed channels.
\Cref{f:missing} shows that there is a significant improvement in prediction performance 
when the number of input users is more than two.
Increasing the number of input users further does not greatly improve \shortname{}'s prediction accuracy. 
This result indicates that \clcp{} is correctly predict channels even when there are many unobserved channels. 

\subsubsection{Overhead Reduction}
We first evaluate runtime distribution of \shortname{} in \Cref{f:runtime}. 
Specifically, \shortname{} achieves only about $4$ ms inference time.
\shortname{}'s inference time comply with that of other VAE-based models~\cite{NEURIPS2020_e3844e18}. 
Next, we present the overhead reduction with varying parameters in \Cref{f:overhead}. 
We define the overhead as the percentage of CSI transmission time over the total traffic time.
The short feedback period, increase in the number of users, and greater number of subcarriers result in a larger CSI overhead in the absence of \shortname{}, making our \clcp{} effective to a greater extent. 
Under densely deployed scenario, our \clcp{} notably reduces the overhead. 
\Cref{f:overhead} (\emph{right}) shows that with $400$ users, \clcp{} can free up more than $40\%$ overhead.

\subsubsection{Channel Capacity.}
In \cref{f:sum_rate}, we evaluate channel capacity of OFDMA packets that are scheduled using predicted channels.
We define channel capacity as a sum of achieved rates at each subcarrier $s$, that is
$R_{\mathrm{capacity}}(RU_{i}) = \sum_{s\in RU_{i}}$ $R_{\mathrm{capacity}}(s)$ where $RU_{i}$ is RU at $i$-th location.
Then, we define
capacity of a complete user schedule $g$ as:
\begin{equation}
\begin{aligned}
    \sum_{j}R_{\mathrm{capacity}}(p_{j},u_{j}) = 
    \sum_{j}\sum_{s\in p_{j}}\sum_{u\in u_{j}}\log_{2}(1+P_{u,s})
\end{aligned}
\label{channel_capacity}
\end{equation}
where $P_{u,s}$ denotes a transmit power for user $u$ and subcarrier $s$. 
In \cref{f:sum_rate}, channel capacity of packets scheduled based on predicted CSIs is almost identical to that of ground-truth CSIs.
These results demonstrate that our predicted channels is accurate enough for OFDMA scheduling.

\subsubsection{PER Distribution}
PER distributions of packets scheduled using predicted CSIs are shown in \cref{f:per}.
Even when PER is high, packets scheduled with ground-truth CSIs and predicted CSIs share similar PER distributions.
We conclude that even if channel condition is bad, \clcp{} still provides accurate channel prediction.

\subsubsection{PHY Rate Distribution}
11ax allows each RU to have its own MCS index, which is calculated based on on its channel condition.
Therefore, rate adaptation requires accurate channel estimates.
In \cref{f:phyrate}, we present PHY rate distributions  
calculated using an effective SNR (ESNR)-based rate adaptation~\cite{halperin2010predictable}.
This algorithm leverages channel estimates to find a proper MCS index.
The results show that PHY rate distributions of both ground-truth channel and predicted channel are highly similar.

\subsubsection{Multiuser Detection}
In 11ax, 
multi-user detection algorithms are used to separate uplink streams from multiple users. 
The drawback is that for uplink MU-MIMO, it is crucial to not only select a subset of 
users with low spatial channel correlation,
but also determine an appropriate decoding precedence.
To evaluate both aspects, we employ several multiuser detection algorithms, such as zero-forcing (ZF) and minimum mean squared error (MMSE), that are integrated with a successive interference cancellation (SIC) technique as well as the most optimal maximum-likelihood (ML) decoder.
\Cref{f:ber} shows a bit-error rate (BER) of packets that are scheduled with ground-truth CSIs and these with predicted CSIs. 
We decode these packets using ZF-SIC, MMSE-SIC, or ML technique across different SNR values. 
We observe that BER of packets from predicted CSIs are slightly higher than packets from ground-truth CSIs for ML decoder when SNR ranges from $10$ to $16$ dB.
On the other hand, BER with ZF- and MMSE-SIC decoder show no difference between predicted and ground-truth CSIs.
This indicates that \shortname{}'s prediction is accurate for ZF-SIC and MMSE-SIC decoders.

\section{Related Work}
\label{s:related}
Work related to CLCP factors into \emph{(i)} work that shares 
some ML techniques with CLCP but which targets other objectives;
and \emph{(ii)} work on predicting \emph{average} wireless
channel strength and the wireless
channel of a single given link, at different frequencies.  
We discuss each in turn
in this section.

\parahead{Deep Probabilistic Networks for Wireless Signals.}
EI~\cite{jiang2018towards} leverages adversarial networks to classify a motion information embedded in the wireless signal. It uses a probabilistic learning model to 
extract environment- and subject-independent features shared by the data collected in different environments. 
RF\hyp{}EATS~\cite{ha2020food} leverages a probabilistic learning framework that adapts the variational inference networks to sense food and liquids in closed containers with the back-scattered RF signals as an input. Like EI, RF\hyp{} builds a model generalized to unseen environments.
Similarly, \shortname{} captures the common information (\textit{e.g.} dynamics in the environment) shared among different wireless links using a deep probabilistic model. 
However, our task is more complicated than removing either the user-specific or environment-specific information. 
We not only decompose observed wireless channels into a representation conveying environment-specific information, but also integrates the representation with user-specific information to generate raw wireless channels of a targeting ''unobserved'' user.

\parahead{Learning-based Channel Prediction.}
A growing body of work leverages various ML techniques for
the broad goals of radio resource management.
CSpy \cite{cspy-mobicom13} uses a Support Vector Machine (SVM)
to predict, on a single link, which channel
out of a set of channels has the strongest average 
magnitude, but does not 
venture into cross-link prediction
at a subcarrier\hyp{}level granularity, which modern
wireless networks 
require in order to perform efficient OFDMA channel allocation
for a group of users.
Also, to manifest compression-based channel sounding for
uplink-dominant massive-IoT networks, 
it requires extremely regular and frequent traffic patterns for every users, which is impractical.
R2F2 \cite{R2F2-sigcomm16} infers downlink CSI for a certain 
LTE mobile\hyp{}base station link based on the path
parameters of uplink CSI readings for that \emph{same} link, 
using an optimization method in lieu of an ML\hyp{}based approach. 
Similarly, \cite{bakshi2019fast} leverages a neural network 
to estimate path parameters, which, in turn, are used to 
predict across frequency bands. 
However, in 802.11ax, 
there is instead a different need and oppportunity: 
to predict \emph{different} links' channels
as recent traffic has used, in order 
to reduce channel estimation overhead, the opportunity \shortname{} targets.

\section{Conclusion}
\label{s:conclusion}

This paper presents the first study to explore cross-link channel prediction 
for the scheduling and resource allocation algorithm in the context of 11ax. 
Our results show that \shortname{}
provides a $2\times$ to $3\times$ throughput gain over baseline and
a $30\%$ to $40\%$ throughput gain over R2F2 and OptML 
in a 144-link testbed. 
To our knowledge, this is the first paper to apply the a 
deep learning-based model to predict channels across links. 

\clearpage
\pagenumbering{roman}
\let\oldbibliography\thebibliography
\renewcommand{\thebibliography}[1]{%
  \oldbibliography{#1}%
  \setlength{\parskip}{0pt}%
  \setlength{\itemsep}{0pt}%
}
\begin{raggedright}
\bibliographystyle{concise2}
\bibliography{reference}
\end{raggedright}

\end{document}